\documentclass[10pt, conference, compsocconf]{IEEEtran}
\usepackage[utf8]{inputenc}

\ifCLASSOPTIONcompsoc
  \usepackage[nocompress]{cite}
\else
  \usepackage{cite}
\fi

\ifCLASSINFOpdf
  \usepackage[pdftex]{graphicx}
  \DeclareGraphicsExtensions{.pdf,.jpeg,.png}
\else
\fi

\ifCLASSOPTIONcompsoc
  \usepackage[caption=false,font=footnotesize,labelfont=sf,textfont=sf]{subfig}
\else
  \usepackage[caption=false,font=footnotesize]{subfig}
\fi

\usepackage[hyphens]{url}

\usepackage{amsmath}
\usepackage{amsthm}
\theoremstyle{plain}
\newtheorem{theorem}{Theorem}

\theoremstyle{definition}
\newtheorem{definition}{Definition}
\newtheorem{example}{Example}

\usepackage{txfonts}

\usepackage{bm}

\usepackage{tikz}
\usetikzlibrary{shapes,arrows,trees,positioning}
\usetikzlibrary{decorations.markings,decorations.pathreplacing,decorations.pathmorphing}
\usetikzlibrary{shapes.geometric, intersections, calc, patterns}
\usetikzlibrary{external}
\usepackage[linesnumbered,ruled,vlined]{algorithm2e}
\SetAlFnt{\small}
\begin{document}

\title{An Incremental Learner for Language-Based Anomaly Detection in XML}
\author{\IEEEauthorblockN{Harald Lampesberger}
\IEEEauthorblockA{Department of Secure Information Systems\\
University of Applied Sciences Upper Austria\\
Email: \url{harald.lampesberger@fh-hagenberg.at}}
}

\maketitle

\begin{abstract}
The Extensible Markup Language (XML) is a complex language, and consequently, XML-based protocols are susceptible to entire classes of implicit and explicit security problems.
Message formats in XML-based protocols are usually specified in XML Schema, and as a first-line defense, schema validation should reject malformed input.
However, extension points in most protocol specifications break validation.
Extension points are wildcards and considered best practice for loose composition, but they also enable an attacker to add unchecked content in a document, e.g., for a signature wrapping attack.

This paper introduces datatyped XML visibly pushdown automata (dXVPAs) as language representation for mixed-content XML and presents an incremental learner that infers a dXVPA from example documents.
The learner generalizes XML types and datatypes in terms of automaton states and transitions, and an inferred dXVPA converges to a good-enough approximation of the true language.
The automaton is free from extension points and capable of stream validation, e.g., as an anomaly detector for XML-based protocols.
For dealing with adversarial training data, two scenarios of poisoning are considered: a poisoning attack is either uncovered at a later time or remains hidden.
Unlearning can therefore remove an identified poisoning attack from a dXVPA, and sanitization trims low-frequent states and transitions to get rid of hidden attacks.
All algorithms have been evaluated in four scenarios, including a web service implemented in Apache Axis2 and Apache Rampart, where attacks have been simulated.
In all scenarios, the learned automaton had zero false positives and outperformed traditional schema validation.
\end{abstract}

\begin{IEEEkeywords}
XML, grammatical inference, visibly pushdown automata, stream validation, anomaly detection, experimental evaluation.
\end{IEEEkeywords}

% \fi
\IEEEpeerreviewmaketitle

\section{Introduction}

The Extensible Markup Language~(XML)~\cite{w3c-xml} is ubiquitous in electronic communication, e.g., the Simple Object Access Protocol~(SOAP), the Extensible Messaging and Presence Protocol~(XMPP), the Security Assertion Markup Language~(SAML), and many data serialization formats.
% ~\cite{w3c-soap} ~\cite{RFC6120} 
The success of XML boils down to its rich data models and tool support: Instead of specifying some protocol from scratch, a software developer can simply define a subset of XML and reuse existing parsing and querying tools.

XML attacks, in particular, the signature wrapping attack~\cite{McIntosh2005}, have motivated this work.
The signature wrapping attack exploits identity constraints and best practices for composition in XML Schema~(XSD)~\cite{w3c-xmlschema}, and the attack's goal is to modify a document without violating a cryptographic signature.
Several high-value targets were vulnerable over the years, e.g., the Amazon EC2 cloud control SOAP interface~\cite{Somorovsky2011} and many SAML frameworks~\cite{Somorovsky2012}.
Attack tool support is already available~\cite{ws-attacker}, and fixing the problem tends to be hard~\cite{Gajek2007,Gajek2009,Jensen2009a,Jensen2011}.
Signature wrapping is a showcase for language-theoretic security because it is the result of design choices.
A document with references is logically not a tree but often wrongly treated as such in modular software, and the need for determinism in composed schemas has led to extension points in many specifications as attack enablers.

This paper extends a previous grammatical inference approach, where a language representation is learned from example documents~\cite{Lampesberger2013b}.
Use cases for the presented approach are anomaly detection in XML-based protocols and schema inference for interface hardening.
The contributions are automaton models as language representations for mixed-content XML, algorithms for datatype inference from texts, an incremental learner, and an experimental evaluation.

For representing event streams of mixed-content XML, the proposed datatyped XML visibly pushdown automata (dXVPAs) and character-data XVPAs (cXVPAs) introduce transitions for text contents in the original XVPA definition by Kumar et al.~\cite{Kumar2007}.
%These models capture event-based XML representations and enable stream validation.
The proposed learner converges to a good-enough language approximation in terms of a dXVPA.
An inferred automaton for stream validation mitigates the signature wrapping attack because it is free from extension points.
Counting the mind changes between incremental steps is a heuristic for measuring convergence in the learning progress.
Furthermore, the learner has been designed with poisoning attacks in mind.
Two scenarios are considered: a successful poisoning attack is uncovered at some later time and a poisoning attack stays hidden but is statistically rare~\cite{Cretu2008}.
For the first case, the learner provides an unlearning operation, and for the second case, a sanitization operation trims low-frequent states and transitions from an automaton.
All algorithms have been implemented and evaluated in four scenarios, where various XML attacks are simulated: two synthetic and two realistic scenarios utilizing Apache Axis2 and Rampart.
In all scenarios, the learned automaton outperformed baseline schema validation.

\subsection{XML}

XML specifies a syntax: open- and close-tags for elements, attributes, namespaces, allowed characters for text content and attribute values, processing instructions for the parser, inline Document Type Definitions (DTD)~\cite{w3c-xml}, and comments.
The syntax allows ambiguities, e.g., an element without text content, and XML Information Set~\cite{w3c-infoset} therefore defines a data model to remove syntactic ambiguities: A document has an infoset if it is \emph{well formed} and all namespace constraints are satisfied.

Business logic accesses infoset items in a document through an interface. 
Common APIs for XML can be distinguished into (a) stream based, e.g., Simple API for XML (SAX) and Streaming API for XML (StAX)~\cite{java-stax} and (b) tree based, e.g., a Document Object Tree (DOM).

A \emph{schema} is basically a grammar, and the XML community provides several schema languages for specifying production rules, e.g., DTD, XSD, Relax NG~\cite{oasis-relaxng}, and Extended DTD~(EDTD)~\cite{Kumar2007} as a generalization.
Productions are of form $a \to B$, where $B$ is a regular expression and called \emph{content model} of $a$.
In DTD, rules are expressed over elements.
To raise expressiveness, productions in XSD, Relax NG, and EDTD are defined over \emph{types}, where every type maps to an element.
This mapping is surjective: two types can map to the same element.

\emph{Schema validation} is checking language acceptance of a document.
\emph{Typing} is stricter than validation by assigning a \emph{unique type} from productions to every element~\cite{Martens2006}.
The power of regular expressions and the surjective relation between types and elements can introduce ambiguity and nondeterminism, but determinism is desired, e.g., for assigning semantics to types.
DTD and XSD therefore have syntactic restrictions to ensure deterministic typing.
Schema validation and typing are first-line defenses against attacks; however, XML identity constraints and extension points in XSD can render validation ineffective.

\subsection{Language-Theoretic Vulnerabilities}

The XML syntax is context free and infoset items are tree structured, but a document is not always logically a tree.
Identity constraints like keys (ID) and key references (IDREF, IDREFS) introduce self-references that go beyond context freeness.
Cyclic and sequential references turn a finite tree data model logically into an infinite tree, and operations such as queries become computationally harder~\cite{Wu2010}.
Furthermore, XSD introduces additional constraints (unique, key, and keyref) over text contents, attribute values, and combinations thereof.
Checking identity constraints during schema validation is costly because indices need to be constructed, or the data model is traversed many times.

Also, there are two philosophies of modularity in XSD: \emph{schema subtyping}~\cite{Lindsay2008} by refining productions and \emph{schema extension points} using wildcards (\texttt{xs:any}).
Extension points allow loose coupling and are considered best practice~\cite{Stephenson2004}.
In an XSD, a wildcard is often accompanied by the \texttt{processContents="lax"} attribute which has a tremendous effect on validation: if there is no schema in the parser's search space for a qualified element at an extension point, validation is skipped.
By choosing a random namespace, an attacker can place arbitrary content at an extension point.
Unfortunately, extension points are present in many standards, e.g., SOAP, XMPP, and SAML.

\subsection{XML Attacks}

Attacks can be distinguished into \emph{parsing} and \emph{semantic} attacks.
Parsing attacks target lexical and syntactical analyses, e.g., for Denial-of-Service.
Examples are oversized tokens, high node counts from unbounded repetitions~\cite{Jensen2009}, and coercive parsing~\cite{wsattacks}.
Schema validation is unable to reject parsing attacks when they are placed at an extension point.
A special class of parsing attacks originates from inline \texttt{DOCTYPE} declarations, i.e., exponential entity expansion, external entities for privilege escalation, and server-side request forgery (SSRF)~\cite{Morgan2014}.

Semantic attacks aim for misinterpretation, e.g., by tampering with structure and texts.
CDATA fields~\cite{wsattacks} can exclude reserved characters (e.g. angled brackets) from lexical analysis as a helper for many semantic attacks, e.g., XML, SQL, LDAP, XPath, and command injection; path traversal; memory corruption in interacting components; and cross-site request forgery (XSRF) and cross-site scripting (XSS) with respect to web applications~\cite{Jensen2009}.

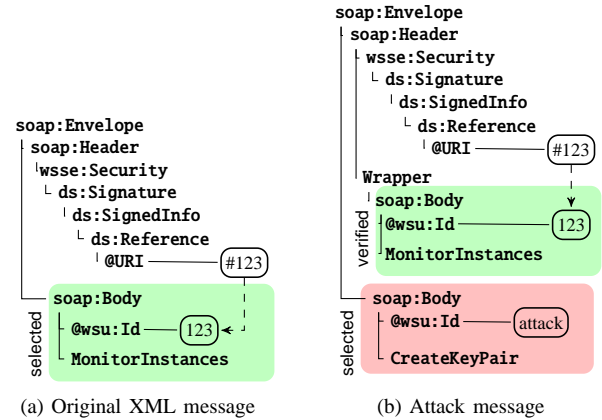
\begin{figure}
	\centering
	\tikzsetnextfilename{signature-wrapping-before}
	\subfloat[Original XML message]{
		\label{fig:signature_wrapping_before}{
		\usetikzlibrary{trees,decorations.pathreplacing,decorations.pathmorphing}
\begin{tikzpicture}[-,>=stealth',shorten >=1pt,auto, font=\sffamily\scriptsize, anchor=base,
  grow via three points={one child at (-0.65,-0.3) and
    two children at (-0.7,-0.3) and (-0.1,-0.3)},
    edge from parent path={(\tikzparentnode.south west) +(0.14,0) |- (\tikzchildnode.west)}]

\tikzstyle{tnode}=[anchor=west, font={\ttfamily\bfseries\scriptsize}, minimum height=1em, inner sep=1.5pt]
\tikzstyle{refnode}=[semithick, font={\scriptsize}, rounded corners, minimum height=1.1em, draw, inner sep=1.8pt]

\draw[fill=green, draw=none, opacity=.25, rounded corners]   (0.4,4) rectangle (3.4,2.7);

\node [tnode] (root) at (-0.1,6.1) {soap:Envelope}
    child { node [tnode] {soap:Header}
        child { node [tnode] {wsse:Security}
        child { node [tnode] {ds:Signature}
            child { node [tnode] {ds:SignedInfo}
              child { node [tnode] {ds:Reference}
                child { node (ref) [tnode] {@URI}}
                }
            }
        }
    }
}
    child { node [tnode] at (-0.3,-2) {soap:Body}
        child { node [tnode] (body1) at ++(0.3,-0.1) {@wsu:Id}}
        child { node [tnode] at (-0.3,-0.5) {MonitorInstances}
         }
    }
;
\node[refnode] (ref0) at (3,4.2) {\#123};
\node[refnode] (ref1) at (2.4,3.3) {123};
\draw (body1) edge (ref1);
\draw (ref) edge (ref0);
\draw[->, dashed]  (ref0) |- (ref1);

%\draw [-,decorate,decoration={brace,amplitude=3pt},xshift=-4pt,yshift=0pt]
%(7.1,3.8) -- (7.1,2.5) node [black,right,midway,xshift=1mm] {Signed};
%\draw [-,decorate,decoration={brace,amplitude=3pt},xshift=-4pt,yshift=0pt]
%(4.4,2.4) -- (4.4,1.1) node [black,right,midway,xshift=1mm,align=left] {Executed by\\Service};

\node[align=center, rotate=90] at (0.3,3.2) {selected};
\end{tikzpicture}
	}}
	\tikzsetnextfilename{signature-wrapping-after}
	\subfloat[Attack message]{
		\label{fig:signature_wrapping_after}{
		\usetikzlibrary{trees,decorations.pathreplacing,decorations.pathmorphing}
\begin{tikzpicture}[-,>=stealth',shorten >=1pt,auto, font=\sffamily\scriptsize, anchor=base,
  grow via three points={one child at (-0.65,-0.3) and
    two children at (-0.7,-0.3) and (-0.1,-0.3)},
    edge from parent path={(\tikzparentnode.south west) +(0.14,0) |- (\tikzchildnode.west)}]

\tikzstyle{tnode}=[anchor=west, font={\ttfamily\bfseries\scriptsize}, minimum height=1em, inner sep=1.5pt]
\tikzstyle{refnode}=[semithick, font={\scriptsize}, rounded corners, minimum height=1.1em, draw, inner sep=1.8pt]

\draw[fill=red, draw=none, opacity=.25, rounded corners]   (0.3,2.5) rectangle (3.4,1.3);
\draw[fill=green, draw=none, opacity=.25, rounded corners]   (0.5,3.8) rectangle (3.5,2.6);

\node [tnode] (root) at (-0.1,6.1) {soap:Envelope}
    child { node [tnode] {soap:Header}
        child { node [tnode] at ++(0.15, 0) {wsse:Security}
        child { node [tnode] {ds:Signature}
            child { node [tnode] {ds:SignedInfo}
              child { node [tnode] {ds:Reference}
                child { node (ref) [tnode] {@URI}}
              }
            }
        }
     }
    child { node [tnode] at ++(-0.5,-1.6) {Wrapper}
      child { node [tnode] at ++(0.3,0) {soap:Body}
        child { node [tnode] at ++(0.2,0) (body1) {@wsu:Id}}
        child { node [tnode] at (-0.4,-0.4) {MonitorInstances}
         }
       }
    }
}
    child { node [tnode] at (-0.3,-3.5) {soap:Body}
        child { node [tnode] (body2) at ++(0.3,-0) {@wsu:Id}}
        child { node [tnode] at (-0.3,-0.5) {CreateKeyPair}
        }
    }
;
\node[refnode] (ref0) at (3.1,4.2) {\#123};
\node[refnode] (ref1) at (3.1,3.2) {123};
\node[refnode] (ref2) at (2.7,1.9) {attack};
\draw (body1) edge (ref1);
\draw (ref) edge (ref0);
\draw (body2) edge (ref2);
\draw[->, dashed]  (ref0) -- (ref1);

%\draw [-,decorate,decoration={brace,amplitude=3pt},xshift=-4pt,yshift=0pt]
%(7.1,3.8) -- (7.1,2.5) node [black,right,midway,xshift=1mm] {Signed};
%\draw [-,decorate,decoration={brace,amplitude=3pt},xshift=-4pt,yshift=0pt]
%(4.4,2.4) -- (4.4,1.1) node [black,right,midway,xshift=1mm,align=left] {Executed by\\Service};

\node[rotate=90] at (0.4,3.1) {verified};
\node[rotate=90] at (0.2,1.7) {selected};
\end{tikzpicture}	
	}}
\caption{XML signature wrapping attack (reproduced from~\cite{Somorovsky2011})}
\label{fig:xml_signature_wrapping}
\end{figure}

\subsubsection{Signature Wrapping Attack}

Signature wrapping is a semantic attack.
XML Signature~\cite{w3c-xmlsignature} specifies a \texttt{ds:signature} element that holds one or more hashes of referenced resources (i.e., elements in the document) and is signed for authenticity.
The resources are referenced by an ID or an XPath expression.
Signature checking verifies the authenticity of the \texttt{ds:signature} element and compares the stored hashes with computed ones.
Checking is usually treated as a Boolean decision, independent from the business logic, and a vulnerability emerges when verified document locations are not communicated between software modules accessing the document.

In a signature wrapping attack, the referenced resource is moved to an extension point, and a malicious element is placed instead at the original location.
An example based on the Amazon EC2 attack~\cite{Somorovsky2011} is shown in Figure~\ref{fig:xml_signature_wrapping}.
As a precondition, the attacker needs access to some signed document (Figure~\ref{fig:signature_wrapping_before}).
The SOAP schema has an extension point in \texttt{soap:Header}, and the original message body is hidden in a wrapper element for skipping schema validation.
The signature remains valid (Figure~\ref{fig:signature_wrapping_after}), and the business logic wrongly processes the attacker-provided message body.

\subsubsection{Signature Wrapping Countermeasures}

\emph{Security policies}~\cite{McIntosh2005} can enforce properties of SOAP messages, but policy checking is computationally costly.
Gruschka and Iacono~\cite{Gruschka2009} furthermore show a successful signature wrapping attack on Amazon EC2 that satisfies security policies.

Rahaman et al.~\cite{Rahaman2006} propose an \emph{inline approach} by adding an element to SOAP headers that stores document characteristics.
Unfortunately, if a single element in the header is not signed, the approach can also be circumvented~\cite{Gajek2007}.

Gajek et al.~\cite{Gajek2007} and Somorovsky et al.~\cite{Somorovsky2012} propose \emph{improved signature verification} by returning a filtered document view, but the business logic needs to be adapted accordingly.
Gajek et al.~\cite{Gajek2009} also propose FastXPath for location-aware XPath-based references in signatures.
Namespace injection in XPath-based references could eventually break this countermeasure too~\cite{Jensen2009a}.

Jensen et al.~\cite{Jensen2011} propose \emph{schema hardening} by removing extension points and restricting repetitions.
Hardening is effective because elements cannot be hidden anymore; however, all composed schemas need to be known beforehand, and generating a single unified hardened schema is computationally hard.
Experiments have also shown a significant slowdown in schema validation.

\subsection{Research Questions}

Removal of extension points is an effective countermeasure, but compiling a unified schema is difficult~\cite{Jensen2011}.
This paper therefore proposes a monitor for an XML-based system.
The monitor has a learner and validator component.
The learner component infers a dXVPA, and the validator component utilizes an optimized variant of the automaton to validate documents sent to the system under observation.
Validation is relative to the training data, and the approach is therefore called \emph{language-based anomaly detection}.
If the validator component rejects a document, some filtering or extended policy checking could be performed, but these operations are not in the scope of this work.

The assumed attacker is capable of reading and modifying documents in transit and sending a malicious document directly to the system under observation.

Clients and services are considered black boxes, where message semantics are unknown to the monitor; however, semantics are important under the language-theoretic security threat model because an attack is basically a misinterpretation.
When a system interprets a document, semantics for elements and texts are derived from assigned types and datatypes respectively, where types and datatypes are usually defined in software (ad hoc) or in schema production rules.
Attacks affect at least one type or datatype in a document for causing misinterpretation.
The system under observation is assumed to have \emph{type-consistent behavior}: for all manifestations of an expected type or datatype in a document, the behavior is well specified.
In other words, language-based anomaly detection only works if attacks are syntactically distinguishable from expected types and datatypes.
To sum up, the research questions are:
\begin{itemize}
	\item [RQ1] What is a suitable language representation for types and datatypes that is capable of stream validation?
	\item [RQ2] Can this language representation be learned?
	\item [RQ3] Can the proposed approach identify attacks?
\end{itemize}

\subsection{Methodology}

\subsubsection{Language Representation}

In mixed-content XML, texts are strings over Unicode, and they are allowed between a start- and an end-tag, an end- and a start-tag, two start-tags, and two end-tags.
XSD provides datatypes for specifying texts, where every datatype has a value space and a lexical space over Unicode.
A language representation that captures document structure and texts needs to be expressive with respect to typing and support \emph{stream validation} for open-ended XML protocols (i.e., XMPP) and very large documents.
To answer RQ1, the paper introduces dXVPAs as an extension of XVPAs~\cite{Kumar2007}.
XVPAs are known to recognize StAX event streams for linear-time stream validation, but text contents are not considered yet.
A dXVPA introduces transitions for datatypes of text content, and a cXVPA is an optimized dXVPA representation for linear-time stream validation in the validator component.

\subsubsection{Learning from Positive Examples}

The learner component receives examples and computes automata for the validation component.
This learning setting corresponds to Gold's \textit{identification in the limit from positive examples}~\cite{Gold1967}, and according to Fernau~\cite{Fernau2003}, the definition is as follows.

\begin{definition}[Identification in the limit from positive examples~\cite{Fernau2003}]
\label{def:limpos}
Let $\mathcal{L}$ be a target language class that can be characterized by a class of language-describing devices~$\mathcal{D}$. 
$E \colon \mathbb{N} \to L$ is an enumeration of strings for a language $L \in \mathcal{L}$, and the examples may be in arbitrary order with possible repetitions.
Target class $\mathcal{L}$ is \emph{identifiable in the limit} if there exists an inductive inference machine or learner $I$:

\begin{itemize}
	\item Learner $I$ receives examples $E(1), E(2), \dots$
	\item Learner $I$ reacts by computing a stream of hypotheses (e.g., automata) $D_1, D_2, \dots$ such that $D_i \in \mathcal{D}$.
	\item For every enumeration of $L \in \mathcal{L}$, there is a convergence point $N(E)$ such that $L = L\left(D_{N(E)}\right)$ and $j \geq N(E) \implies D_j = D_{N(E)}$.
\end{itemize}
\end{definition}

RQ2 is answered by specifying algorithms for inferring datatypes from text and automata from documents.
Furthermore, unlearning and sanitization operations for dealing with adversarial training data are provided.

\subsubsection{Experimental Evaluation}

A learning-based approach is still a heuristic and requires experimental evaluation.
Four datasets have been generated: two synthetic ones using a stochastic XML generator and two realistic ones from a web service implemented in Apache Axis2 and Apache Rampart.
The service has been implemented according to best practices, and attacks have been performed manually and automatically by the WS-Attacker~\cite{ws-attacker} tool.
Detection performance, learning progress in terms of mind changes, and the effects of unlearning and sanitization have been analyzed to answer RQ3.

\section{Grammatical Inference of XML}

\tikzsetnextfilename{incremental-algorithms}
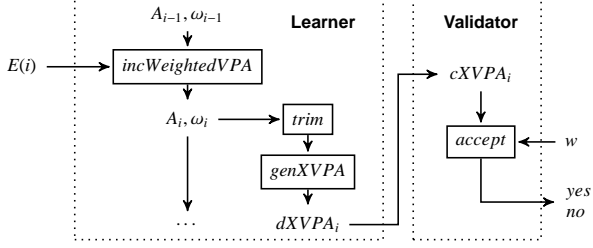
\begin{figure}
\centering
\begin{tikzpicture}[->,>=stealth',shorten >=1pt,semithick, font=\sffamily\scriptsize]
\tikzstyle{alg} = [draw]

\node (v1) at (0.5,0.1) {$E(i)$};
\node (v2) at (2.7,-0.6) {$A_i, \omega_i$};
\draw[dotted]  (1.2,1) rectangle (5.3,-2.2);
\draw[dotted]  (5.7,1) rectangle (7.4,-2.2);
\node (v11) at (4.3,-2) {$dXVPA_i$};
\node (v16) at (6.6,0) {$cXVPA_i$};
\node at (4.5,0.7) {\textsf{\textbf{Learner}}};
\node at (6.6,0.7) {\textsf{\textbf{Validator}}};
\node[alg] (v5) at (2.7,0.1) {$incWeightedVPA$};
\node[alg] (v13) at (4.3,-1.3) {$genXVPA$};
\node[alg] (v20) at (6.6,-0.9) {$accept$};
%\node (v6) at (2.4,1) {$VPA_0$};
\draw  (v1) edge (v5);
\node (v6) at (7.8,-0.9) {$w$};
\node[align=left] (v23) at (7.9,-1.7) {$yes$\\$no$};
\node (v3) at (2.7,0.8) {$A_{i-1}, \omega_{i-1}$};
\draw  (v3) edge (v5);
\draw  (v5) edge (v2);
\node[alg] (v4) at (4.3,-0.6) {$trim$};
%\node[alg] (v7) at (4.4,-2.6) {$minimize$};
\draw  (v2) edge (v4);
\draw  (v4) edge (v13);
%\draw[<->]  (v13) -- (v7);
\draw  (v13) edge (v11);
\draw  (v11) -- ++(1.2,0) |- (v16);
\draw  (v20) |-(v23);
\draw  (v16) edge (v20);
\draw  (v6) edge (v20);
\node (v8) at (2.7,-2) {$\dots$};
\draw  (v2) edge (v8);
\end{tikzpicture}
\caption{Incremental learning step}
\label{fig:incremental_algorithms}
\end{figure}

Figure~\ref{fig:incremental_algorithms} illustrates the incremental learning step.
The learner component maintains an internal visibly pushdown automaton (VPA) $A$ and counters $\omega_\delta, \omega_Q, \omega_F$ for transitions, states, and final states.
A VPA is a special pushdown automaton with three disjoints alphabets: a call alphabet that pushes on the stack, an internal alphabet that leaves the stack unchanged, and a return alphabet that pops from the stack.
This concept originates from program analysis, and for XML, the alphabets represent different kinds of events.
The set of states is implicitly the stack alphabet.
For a complete definition of VPAs, the reader is directed to Alur et al.~\cite{Alur2004}.

Algorithm~\ref{alg:incwvpa} ($incWeightedVPA$) receives a document event stream $E(i)$ and updates the VPA and the counters.
The counters are frequencies of states and transitions from training data and necessary for unlearning and sanitization operations.
Algorithm~\ref{alg:trim} ($trim$) removes zero-weight states and transitions, and Algorithm~\ref{alg:genxvpa} ($genXVPA$) constructs a minimized dXVPA.
The dXVPA becomes an optimized cXVPA for the validator component, and acceptance of documents can then be efficiently decided.

\subsection{Document Event Stream}

\begin{definition}[Document event stream]
\label{def:event_stream}
A \emph{document event stream} $w$ is a sequence of StAX events $e$ carrying values $lab(e)$.
There are three kinds of events: $startElement$ and $endElement$ for open- and close-tags of qualified element names and $characters$ for texts.
Processing instructions, comments, and entity references are ignored.
Attributes are alphabetically sorted, treated as elements with a leading \texttt{@} symbol, and mapped to a subsequence of $startElement$, $characters$, and $endElement$ events.

\end{definition}

For simpler notation, a $startElement$ event for qualified element $\mathtt{m}$ is denoted as $\mathtt{m}$, and $\overline{\mathtt{m}}$ is the respective $endElement$ event.
The value of a $characters$ event is a string over Unicode, and nested CDATA sections are automatically unwrapped by the parser.

XSD provides datatypes for specifying text contents.
In this work, only the lexical spaces of XSD datatypes~\cite{w3c-xsd-datatypes} are considered in a generalized notation of lexical datatypes.

\begin{definition}[Lexical datatypes]\label{def:datatypes}
Let $T$ be a set of \emph{lexical datatypes}.
A lexical space is a regular language over Unicode $U$, and $\phi : T \to REG(U)$ assigns lexical spaces.
\end{definition}

Lexical datatypes allow to define datatyped event streams, where datatypes replace text contents in $characters$ events.

\begin{definition}[Datatyped event stream]
A \emph{datatyped event stream} $w'$ is a sequence of $startElement$, $endElement$, and $characters$ events.
The value of a $characters$ event $e$ is a datatype $lab(e) \in T$.
A document event stream $w$ \emph{corresponds} to a datatyped event stream $w'$ if $w$ and $w'$ have congruent event kinds, the qualified element names in $startElement$ and $endElement$ events are the same, and text content in a $characters$ event in $w$ is in the lexical space of the congruent $characters$ event in $w'$.
\end{definition}

\subsection{Language Representation}

\tikzsetnextfilename{dxvpa-cardealer}
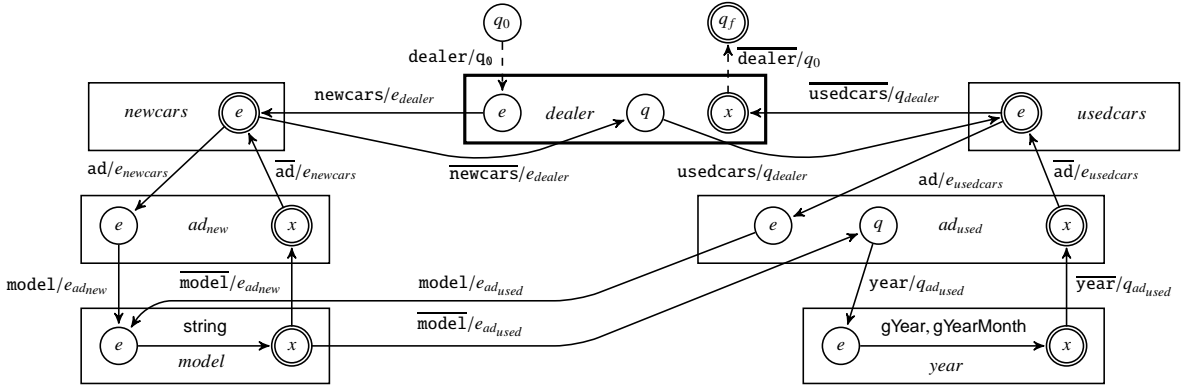
\begin{figure*}
\centering
\begin{tikzpicture}[->,>=stealth',shorten >=1pt,semithick]
\tikzstyle{sm2} = [circle, draw, inner sep=0pt,minimum size=14pt]
\tikzstyle{every node}=[font=\scriptsize]
\node[sm2] (v1) at (-3.5,1) {$e$};
\node[sm2] (v6) at (-1.6,1) {$q$};
\node[sm2, double] (v11) at (-0.5,1) {$x$};
\draw[very thick]  (-4,1.5) rectangle (0,0.6);
\node at (-2.6,1) {${dealer}$};
\node[sm2, double] (v2) at (-7,1) {$e$};
\draw  (-6.4,0.5) rectangle (-9,1.4);
\node at (-8.1,1) {${newcars}$};
\node[sm2,double] (v10) at (3.4,1) {$e$};
\draw  (5.5,0.5) rectangle (2.7,1.4);
\node at (4.6,1) {${usedcars}$};
\node[sm2] (v3) at (-8.6,-0.5) {$e$};
\node[sm2, double] (v5) at (-6.3,-0.5) {$x$};
\draw  (-5.8,-1) rectangle (-9.1,-0.1);
\node at (-7.4,-0.5) {${ad_{new}}$};
\node[sm2] (v7) at (0.1,-0.5) {$e$};
\node[sm2] (v12) at (1.5,-0.5) {$q$};
\node[sm2,double] (v9) at (4,-0.5) {$x$};
\draw  (4.5,-1) rectangle (-0.9,-0.1);
\node at (2.6,-0.5) {${ad_{used}}$};
\node[sm2] (v4) at (-8.6,-2.1) {$e$};
\draw  (-5.8,-2.6) rectangle (-9.1,-1.6);
\node at (-7.5,-2.3) {${model}$};
\node[sm2] (v8) at (1,-2.1) {$e$};
\draw  (4.5,-2.6) rectangle (0.5,-1.6);
\node at (2.4,-2.4) {${year}$};
\node[sm2, double] (v15) at (-6.3,-2.1) {$x$};
\node[sm2, double] (v16) at (4,-2.1) {$x$};
\path (v1) edge[bend right=0, above, pos=0.48] node {$\mathtt{newcars} / e_{dealer}$} (v2)
(v2) edge[bend left=0, left] node {$\mathtt{ad} / e_{newcars}$} (v3)
(v3) edge[bend left=0, left] node {$\mathtt{model} / e_{ad_{new}}$} (v4)
(v15) edge node[left, pos=0.55] {$\overline{\mathtt{model}} / e_{ad_{new}}$} (v5)
(v5) edge[bend right=0, right] node {$\overline{\mathtt{ad}} / e_{newcars}$} (v2)
(v10) edge[bend right=0, below right, pos=0.4] node {} (v7)
(v12) edge[bend left=0, right, pos=0.5] node {$\mathtt{year} / q_{ad_{used}}$} (v8)
(v16) edge[bend left=0, right, pos=0.5] node {$\overline{\mathtt{year}} / q_{ad_{used}}$} (v9)
(v9) edge[bend left=0, right, pos=0.5] node {$\overline{\mathtt{ad}} / e_{usedcars}$} (v10)
(v10) edge[bend right=0, above, pos=0.5] node {$\overline{\mathtt{usedcars}} / q_{dealer}$} (v11);
\node[sm2] (v13) at (-3.5,2.2) {$q_0$};
\node[sm2, double] (v14) at (-0.5,2.2) {$q_f$};
\draw[dashed]  (v13) edge node [left, pos=0.3] {$\mathtt{dealer/q_0}$} (v1);
\draw[dashed]  (v11) edge node [right, pos=0.7] {$\overline{\mathtt{dealer}}/q_0$} (v14);
\draw[rounded corners=3mm] (v15) -- node[above, pos=0.6] {$\overline{\mathtt{model}} / e_{ad_{used}}$}  ++(3.8,0) -- (v12);
\draw  (v4) edge node [above] {$\mathsf{string}$} (v15);
\draw[rounded corners=3mm] (v7) -- ++(-2.6,-1) -- ++(-5.7,0) -- (v4);
\draw[rounded corners=7mm] (v6)  -- ++(1.9,-0.7) --  (v10);
\node at (-3.9,-1.3) {$\mathtt{model} / e_{ad_{used}}$};
\draw  (v8) edge node [above] {$\mathsf{gYear, gYearMonth}$} (v16);
\node at (-3.4,0.2) {$\overline{\mathtt{newcars}} / e_{dealer}$};
\node at (-0.3,0.2) {$\mathtt{\mathtt{usedcars}} / q_{dealer}$};
\node at (2.6,0.1) {$\mathtt{ad} / e_{usedcars}$};
\draw[rounded corners=7mm] (v2) -- ++(3.4,-0.7) -- (v6);
\end{tikzpicture}
\caption{A dXVPA example}
\label{fig:cardealer_example}
\end{figure*}

\subsubsection{Datatyped XVPA}

The dXVPAs extend XVPAs~\cite{Kumar2007} with datatypes, so they can accept datatyped event streams.

\begin{definition}[dXVPA]
A \emph{dXVPA} $A$ over $(\Sigma, M, \mu, T, \phi)$ is a tuple $A = (\{Q_m, e_m, X_m, \delta_m\}_{m \in M}, m_0, F)$.
$\Sigma$ is a set of qualified element names, $M$ is a set of modules (equivalent to types in schemas), $\mu: M \to \Sigma$ is a surjective mapping that assigns elements to modules, $T$ is a set of datatypes, and $\phi: T \to REG(U)$ assigns lexical spaces over Unicode.

For every module $m \in M$:
\begin{itemize}
	\item $Q_m$ is a finite set of module states
	\item $e_m \in Q_m$ is the module's single entry state
	\item $X_m \subseteq Q_m$ are the module's exit states
	\item $\delta_m = \delta_m^{call} \uplus \delta_m^{int} \uplus \delta_m^{ret}$ are module transitions
	\begin{itemize}
		\item $\delta_m^{call} \subseteq \{q_m \xrightarrow{\mathtt{c}/q_m} e_n \mid n \in \mu^{-1}(\mathtt{c})\}$, where $\mathtt{c}$ is a $startElement$ event value that pushes $q_m$ on the stack
		\item $\delta_m^{int} \subseteq \{q_m \xrightarrow{\tau} p_m \mid \tau \in T \}$ and $\tau$ is the value of a datatyped $characters$ event
		\item $\delta_m^{ret} \subseteq \{q_m \xrightarrow{\overline{\mathtt{c}}/p_n} q_n \mid n \in \mu^{-1}(\mathtt{c})\}$, where $\overline{\mathtt{c}}$ is an $endElement$ event value that pops $p_n$ from the stack; the relation is deterministic, i.e., $q_n = q_n'$ whenever $q_m \xrightarrow{\overline{\mathtt{c}}/p_n} q_n$ and $q_m \xrightarrow{\overline{\mathtt{c}}/p_n} q_n'$
	\end{itemize}
\end{itemize}

Module $m_0 \in M$ is the start module, $F = X_{m_0}$ are final states, automaton $A$ satisfies the \emph{single-exit property}~\cite{Kumar2007}, and all transitions satisfy \emph{mixed-content restrictions}.
\end{definition}

The module states are implicitly a stack alphabet.
Call transitions save the current state on the stack and move to the entry of a module.
Internal transitions leave the stack unchanged.
Return transitions pop the stack and move from an exit state to a state in the previous module.

The \emph{single-exit property} ensures that every exit state in a module has the same return transitions.
Otherwise, the exit states would behave differently depending on the saved state on top of the stack.
In XML, modules represent types, and the language of a type must be independent from the parent type, i.e., the calling module.

\begin{definition}[Mixed-content restrictions]\label{def:mixed}
Datatypes are not allowed to affect typing of elements, and the two \emph{mixed-content restrictions} must be satisfied:
\begin{itemize}
	\item \emph{Datatype choice.} \ A datatype choice at state $q_m$ must lead to a single next state, i.e., if $q_m \xrightarrow{\tau} q'_m \in \delta_m^{int}$ and $q_m \xrightarrow{\tau'} q''_m \in \delta_m^{int}$ then $q'_m = q''_m$.
	\item \emph{Datatype sequence.} \ A return transition can only move to a state that is not a successor of a datatype choice, i.e., if $\exists q. q \xrightarrow{\overline{\mathtt{n}}/p_m} q_m \in \delta_n^{ret}$ then $\forall q'. q' \xrightarrow{\tau} q_m \notin \delta_m^{int}$.
\end{itemize}
\end{definition}

The restrictions guarantee that after a $characters$ event, a module is either exited or another module is called because there can never be two subsequent $characters$ events in our definition.

The semantics of dXVPA $A$ are characterized by VPA $A' = (Q, q_0, \{q_f\}, Q, \delta)$ over the visibly pushdown alphabet $(\Sigma \uplus T \uplus \overline{\Sigma})$ by introducing a start state $q_0$ and final state $q_f$:

\begin{align*}
Q &= \{q_0, q_f\} \cup \bigcup_{m \in M} Q_m  \\
\delta &= \{q_0 \xrightarrow{\mu(m_0)/q_0} e_{m_0}\} \cup \{q \xrightarrow{\overline{\mu(m_0)}/q_0} q_f \mid q \in F\} \cup \bigcup_{m \in M} \delta_m
\end{align*}

A run of $A'$ is denoted as $(q_0, \bot) \xrightarrow{w}_A (q, v)$, where $w$ is a datatyped event stream, $q$ is the reached state, and $v$ is the stack.
Event stream $w$ is accepted if $q = q_f$ and $v = \bot$.
Automaton $A$ accepts language $L(A)= L(A')$.

Kumar et al.~\cite{Kumar2007} have also shown that for every EDTD an XVPA that accepts the same language can be constructed and vice-versa.
This theorem can be extended to dXVPAs, but this exceeds the scope of this paper.

\begin{example}
Consider the following EDTD schema.
The qualified elements are $\Sigma = \{$\texttt{dealer}, \texttt{newcars}, \texttt{usedcars}, \texttt{ad}, \texttt{model}, \texttt{year}$\}$, the types are $M=\{dealer,$ $newcars,$ $usedcars,$ $ad_{new},$ $ad_{used},$ $model,$ $year\}$, the start type is $dealer$, and productions over types are:

\vspace{-1.3em}

\begin{align*}
	& d(dealer) \mapsto newcars \cdot oldcars & & d(newcars) \mapsto ad_{new}^*  \\
	& d(usedcars) \mapsto ad_{used}^* & & d(ad_{new}) \mapsto model  \\
	& d(ad_{used}) \mapsto model \cdot year & & d(model) \mapsto \mathsf{string} \\
	& d(year) \mapsto \mathsf{gYear + gYearMonth} \\
\end{align*}

\vspace{-1.3em}

In XSD jargon, type $model$ and $year$ are simple types, and the others are complex types.
Note that element \texttt{ad} has a different type depending on its context in a document.
Figure~\ref{fig:cardealer_example} illustrates the equivalent dXVPA, where states $q_0$ and $q_f$ are added to highlight the VPA semantics.
The dXVPA modules are the types.
Module ${model}$ is called by modules ${ad}_{new}$ and ${ad}_{used}$, and runs return correctly based on the saved stack value.
\end{example}

\subsubsection{Character-Data XVPA}

A dXVPA cannot validate document event streams efficiently.
If a dXVPA is in state $p_m$ and a $characters$ event $e$ encountered, the automaton can only proceed to some state $q_m$ if there is an internal transition $p_m \xrightarrow{\tau} q_m \in \delta_m^{int}$ and the event's text content $lab(e)$ is in the lexical space of the datatype $lab(e) \in \phi(\tau)$.
In the worst case, $lab(e)$ needs to be buffered and checked for every possible datatype.
A cXVPA unifies a datatype choice between two states into a predicate $\psi \in \Psi$, so a text needs to be checked only once during validation.

\begin{definition}[cXVPA]
A \emph{cXVPA} $A$ over $(\Sigma, M, \mu, \Psi)$ is a tuple $A = (\{Q_m, e_m, X_m, \delta_m\}_{m \in M}, m_0, F)$ and adapts the dXVPA definition by using $\delta_m^{int}: Q_m \times \Psi \to Q_m$ for internal transitions.
At most one internal transition per state is allowed, i.e., if $p_m \xrightarrow{\psi_i} q_m$ and $p_m \xrightarrow{\psi_j} q'_m$ then $q_m = q'_m$ and $\psi_i = \psi_j$.
\end{definition}

Same as for dXVPAs, the semantics and accepted language of a cXVPA are given by the corresponding VPA over $(\Sigma \uplus \Psi \uplus \overline{\Sigma})$, where $\Psi$ are predicates over Unicode strings.
A run on an event stream moves along an internal transition $p_m \xrightarrow{\psi} q_m \in \delta^{int}_m$ if $\psi(lab(e))$ holds in state $p_m$.

\begin{theorem}
\label{thm:cxvpa}
Every dXVPA has an equivalent cXVPA for efficiently checking acceptance of document event streams.
\end{theorem}

To sketch the proof, the mixed-content restrictions in dXVPAs enforce that at most one successor state is reachable through internal transitions.
A set of internal transitions is replaced by a single predicate transition.
Lexical spaces in Definition~\ref{def:datatypes} are regular languages, where union is closed.
A unified deterministic finite automaton (DFA) is constructed to represent a predicate, and acceptance of a text can then be decided in a single pass and linear time.

\subsection{Datatype Inference from Text Content}

Given some datatype, it is straightforward to check if texts are within the datatype's lexical space.
But a learner only observes texts without any datatype information.
For a first generalization from texts to datatypes, a lexical datatype system is therefore proposed.
The lexical datatype system infers a set of \emph{minimally required datatypes} for a text content by lexical subsumption and a preference heuristic.

\begin{definition}[Lexical datatype system]
A \emph{lexical datatype system} is a tuple $dts = (T, \phi, \sim_s, \leq_s)$, where $T$ and $\phi$ are according to Definition~\ref{def:datatypes}.
Datatypes must be lexically distinct, i.e., $\phi(\tau) = \phi(\tau') \implies \tau = \tau'$, and $\phi$ imposes a partial ordering $\tau \leq_{lex} \tau' \iff \phi(\tau) \subseteq \phi(\tau')$.
With respect to $\leq_{lex}$, $T$ always contains a unique top datatype $\top$ that accepts any string, i.e., $\phi(\top) \mapsto U^*$.
Equivalence relation $\sim_s: T \to K$ partitions datatypes into semantically related kinds $K$, and $\leq_s$ is an ordering on kinds.
Moreover, the kinds impose a semantic ordering $\leq'_s$ on datatypes, i.e., $\tau \leq'_s \tau' \iff [\tau]_{\sim_s} \leq_s [\tau']_{\sim_s}$.
\end{definition}

\tikzsetnextfilename{lexical-datatype-system}
\begin{figure}
\centering
\begin{tikzpicture}[->,>=stealth',shorten >=1pt,auto,semithick]
\tikzstyle{every node}=[font=\scriptsize\sffamily, anchor=base]
%\node (v1) at (-4.4,1) {booleanNum};
\node (v3) at (-4.1,6) {{boolean}};
\node (v2) at (-9.5,2.8) {{unsignedByte}};
\node (v4) at (-10.7,2.8) {{byte}};
\node (v5) at (-5.3,5.2) {{language}};
\node (v6) at (-4.1,7.7) {{NCName}};
\node (v7) at (-6.5,5.2) {duration};
\node (v9) at (-5.2,4.4) {{dayTimeDuration}};
\node (v8) at (-7.4,4.4) {{yearMonthDuration}};
%\draw  (v1) edge (v2);
%\draw  (v1) edge (v3);
%\draw  (v1) edge (v4);
%\draw  (v3) edge (v5);
% \draw  (v5) edge (v6);
% \draw  (v7) edge (v6);
% \draw  (v8) edge (v7);
% \draw  (v9) edge (v7);
\node (v10) at (-5.3,8.4) {QName};
\node (v11) at (-5.3,6) {Name};
\node (v12) at (-5.9,7.7) {NMTOKEN};
\node (v13) at (-6.7,9.2) {token};
\node (v14) at (-7.7,9.9) {normalizedString};
\node (v15) at (-7.7,10.7) {string};
\node (top) at (-7.7,11.4) {$\top$};
\draw (v15) edge (top);
% \draw  (v6) edge (v10);
% \draw  (v10) edge (v11);
% \draw  (v11) edge (v12);
% \draw  (v12) edge (v13);
% \draw  (v13) edge (v14);
% \draw  (v14) edge (v15);
%\node (v17) at (-3.5,12.1) {base64BinaryLF};
\node (v16) at (-5.6,9.9) {{base64Binary}};
%\draw  (v16) edge (v17);
%\draw  (v17) edge (v15);
\draw  (v16) edge (v15);
\node (v18) at (-10.7,10.7) {{gMonth}};
\node (v19) at (-10.7,11) {{gDay}};
\node (v20) at (-10.7,10.1) {{gMonthDay}};
\node (v21) at (-10.7,9.8) {{gYearMonth}};
% \draw  (v18) edge (v12);
% \draw  (v19) edge (v12);
% \draw  (v20) edge (v12);
% \draw  (v21) edge (v12);
\node (v24) at (-9.1,7.7) {double};
\node (v23) at (-10.7,7.7) {decimal};
\node (v22) at (-10.7,7) {integer};
% \draw  (v22) edge (v23);
% \draw  (v23) edge (v24);
% \draw  (v24) edge (v12);
\node (v25) at (-9.5,3.6) {unsignedShort};
\node (v26) at (-9.5,4.4) {unsignedInt};
\node (v27) at (-7.8,5.2) {unsignedLong};
\node (v28) at (-9.5,6) {nonNegativeInteger};
% \draw  (v2) edge (v25);
% \draw  (v25) edge (v26);
% \draw  (v26) edge (v27);
% \draw  (v27) edge (v28);
% \draw  (v28) edge (v22);
\node (v29) at (-10.7,3.6) {short};
\node (v30) at (-10.7,4.4) {int};
\node (v31) at (-10.7,5.2) {long};
% \draw  (v4) edge (v29);
% \draw  (v29) edge (v30);
% \draw  (v30) edge (v31);
% \draw  (v31) edge (v22);
\node (v34) at (-10.7,10.4) {gYear};
\node (v33) at (-8.7,7) {nonPositiveInteger};
\node (v32) at (-8,6.3) {{negativeInteger}};
% \draw  (v32) edge (v33);
% \draw  (v33) edge (v22);
% \draw  (v34) edge (v22);
%\node (v35) at (-6.9,5.1) {evenLenInteger};
% \draw  (v35) edge (v28);
\node (v36) at (-4.1,6.7) {{hexBinary}};
% \draw  (v35) edge (v36);
% \draw  (v36) edge (v6);
\node (v41) at (-7.7,7.7) {anyURI};
\node (v38) at (-10.7,8.9) {dateTime};
\node (v39) at (-10.7,8.2) {{dateTimeStamp}};
\node (v40) at (-10.7,9.2) {{time}};
\node (v37) at (-10.7,9.5) {{date}};
% \draw  (v37) edge (v38);
% \draw  (v39) edge (v38);
% \draw  (v40) edge (v38);
% \draw  (v41) edge (v12);
% \draw  (v10) edge (v41);
% \draw  (v38) edge (v12);
% \draw  (v2) edge (v29);
% \draw  (v25) edge (v30);
% \draw  (v26) edge (v31);
%\node (v44) at (-2.5,0.1) {boolean0};
%\node (v42) at (-5.7,0.1) {boolean1};
\node (v43) at (-9.5,5.2) {{positiveInteger}};
%\draw  (v42) edge (v43);
% \draw  (v43) edge (v28);
%\draw  (v44) edge (v33);
%\draw  (v44) edge (v1);
%\draw  (v42) edge (v1);
\draw  (v14) edge (v15);
\draw  (v13) edge (v14);
\node (v1) at (-6.7,8.4) {NMTOKENS};
\draw  (v12) edge (v1);
\draw  (v1) edge (v13);
\node (v17) at (-4.1,8.4) {ENTITIES};
\draw  (v6) edge (v17);
\draw  (v6) edge (v10);
\draw  (v17) edge (v13);
\draw  (v10) edge (v13);
\draw  (v24) edge (v13);
\draw  (v21)  -| ++(1.2,0.1) |- (v13);
\draw  (v37)  -| ++(1.2,0.1) |- (v13);
\draw  (v38) -| ++(1.2,0.1) |- (v13);
\draw  (v34)  -| ++(1.2,0.1) |- (v13);
\draw  (v39) edge (v38);
\draw  (v20)  -| ++(1.2,0.1) |- (v13);
\draw  (v18)  -| ++(1.2,0.1) |- (v13);
\draw  (v40)  -| ++(1.2,0.1) |- (v13);
\draw  (v19)  -| ++(1.2,-0.1) |- (v13);
\draw  (v41) edge (v14);
\draw  (v11) edge (v12);
\draw  (v24) edge (v41);
\draw  (v23) edge (v24);
\draw  (v22) edge (v23);
\draw  (v36) edge (v41);
\draw  (v3) edge (v41);
\draw  (v3) edge (v12);
\draw  (v36) edge (v12);
\draw  (v7) edge (v41);
\draw  (v7) edge (v12);
\draw  (v8) edge (v7);
\draw  (v9) edge (v7);
\draw  (v5) edge (v11);
\draw  (v27) edge (v12);
\draw  (v32) edge (v12);
\draw  (v32) edge (v33);
\draw  (v33) edge (v22);
\draw  (v2) edge (v25);
\draw  (v25) edge (v26);
\draw  (v26) edge (v27);
\draw  (v27) edge (v28);
\draw  (v28) edge (v22);
\draw  (v31) edge (v22);
\draw  (v4) edge (v29);
\draw  (v29) edge (v30);
\draw  (v30) edge (v31);
\draw  (v2) edge (v29);
\draw  (v25) edge (v30);
\draw  (v26) edge (v31);
\draw  (v43) edge (v28);
\draw  (v5) edge (v41);
\end{tikzpicture}
\caption{Ordering $\leq_{lex}$ on lexically distinct XSD datatypes}
\label{fig:lexical_datatype_system}
\end{figure}

\subsubsection{Lexical Subsumption}

Figure~\ref{fig:lexical_datatype_system} illustrates the datatype system based on primitive and build-in XSD datatypes~\cite{w3c-xsd-datatypes}.
The standard specifies lexical spaces of datatypes as Unicode regular expressions, and $\leq_{lex}$ is computed from those specifications.
Some datatypes are lexically indistinguishable and are therefore not included: $\textsf{double} =_{lex} \textsf{float}$, $\textsf{NCName} =_{lex} \textsf{ENTITY} =_{lex} \textsf{ID} =_{lex} \textsf{IDREF}$, and $\textsf{NMTOKENS} =_{lex} \textsf{ENTITIES} =_{lex} \textsf{IDREFS}$.
For text contents, a learner needs to infer the least lexical space approximated by a set of datatypes (i.e., a datatype choice).

\begin{definition}[Minimally required datatypes]
The set of \emph{minimally required datatypes} for a Unicode string $w$ is the nonempty antichain $R \subseteq T$ of minimal datatypes with respect to $\leq_{lex}$ such that $\tau \in R \implies w \in \phi(\tau)$, and $\tau' <_{lex} \tau \implies w \notin \phi(\tau')$.
\end{definition}

\begin{algorithm}
\caption{$minLex$}
\label{alg:minlex}
  \DontPrintSemicolon
  \SetKw{In}{in}
  \SetKw{And}{and}
  \SetKwInOut{Input}{Input}
  \Input{lexical datatype system $(T, \phi, \sim_s, \leq_s)$\\Unicode string $w$}
  \KwOut{minimally required datatypes $R \subseteq T$}
  \BlankLine
  $R := \emptyset$; $cand := T$ \;
  \For{$\tau$ \In $topologicalSortOrder(T, \leq_{lex})$} {
  	\lIf{$cand = \emptyset$}{done}
  	\ElseIf{$\tau \in cand$ \And $w \in \phi(\tau)$} {
  		$R:= R \cup \{\tau\}$ \;
  		$cand:= cand \; \setminus \; \uparrow\tau$ w.r.t. $\leq_{lex}$
  	}
  }
\end{algorithm}

The minimally required datatypes are computed by in Algorithm~\ref{alg:minlex} ($minLex$).
The algorithm terminates after $|T|$ steps in the worst case.
Acceptance of a string by a datatype is checked in topological sort order with respect to $\leq_{lex}$.
To minimize the number of checks, a candidates set $cand$ is maintained.
If $w$ is in some lexical space, $w$ is also in all greater datatypes because $\leq_{lex}$ is transitive, and the up-set can be removed from $cand$.
Furthermore, the topological order guarantees that the matched datatypes are minimal and incomparable.
Algorithm $minLex$ always returns a nonempty set because the $\top$ datatype has space $U^*$ and matches for any string.

\tikzsetnextfilename{dts-preference-heuristic}
\begin{figure}
\centering
\begin{tikzpicture}[->,>=stealth',shorten >=1pt,auto,semithick]
\tikzstyle{every node}=[font=\scriptsize\sffamily, anchor=base]

\node (v13) at (3.2,13) {encodingLike};
\node (v14) at (1.2,14.4) {structureLike};
\node (v15) at (3.2,14.4) {stringLike};
\node (top) at (3.2,15.1) {$\top$};
\draw (v15) edge (top);
\node (v16) at (1.2,10.8) {{booleanLike}};
\node (v18) at (1.2,12.3) {atomicNumericLike};
\node (v24) at (1.2,11.6) {atomicUnsignedLike};
\node (v23) at (3.2,10.8) {{listLike}};
\node (v34) at (1.2,13) {numericLike};
\node (v38) at (1.2,13.7) {temporalLike};

\draw  (v13) edge (v15);

\draw  (v38) edge (v14);
\draw  (v18) edge (v34);
\draw  (v24) edge (v18);

\draw  (v34) edge (v13);
\draw  (v34) edge (v38);

\draw  (v23) edge (v13);
\draw  (v14) edge (v15);
\draw  (v16) edge (v24);

\end{tikzpicture}
\caption{Ordering $\leq_{s}$ on kinds of lexical datatypes}
\label{fig:preference_heuristic}
\end{figure}
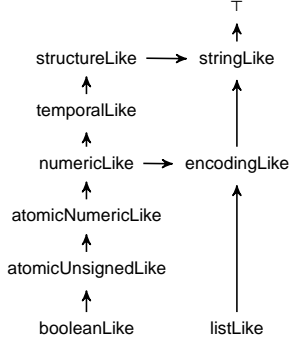

\subsubsection{Preference Heuristic}

Figure~\ref{fig:lexical_datatype_system} already suggests that lexical spaces of XSD datatypes are often incomparable and ambiguous.
This leads to weird datatype choices, e.g., $minLex(\texttt{false})=\{\textsf{language, boolean, NCName}\}$.
The antichain is lexically correct, but some datatypes are semantically more informative and preferred over others.
A second step in datatype inference is therefore to drop the least informative datatypes from minimally required datatypes.
The proposed heuristic captures the XSD type hierarchy and datatype semantics in an ordering $\leq_s$ for kinds of datatypes.
Figure~\ref{fig:preference_heuristic} illustrates the ordering, and kinds are defined as:

{\scriptsize
\begin{align*}
&\textsf{stringLike} = \{\textsf{string, normalizedString, token, ENTITY, ID,IDREF, NMTOKEN}\} \\
&\textsf{listLike} = \{\textsf{ENTITIES, IDREFS, NMTOKENS}\} \\
&\textsf{structureLike} = \{\textsf{anyURI, NOTATION, QName, Name, language, NCName}\} \\
&\textsf{encodingLike} = \{\textsf{base64Binary, hexBinary}\} \\
&\textsf{temporalLike} = \{\textsf{gDay, gMonth, gYear, gYearMonth, gMonthDay, date, duration,} \\
&\phantom{{}=1}\textsf{time, dayTimeDuration, yearMonthDuration, dateTime, dateTimeStamp}\} \notag \\
&\textsf{numericLike} = \{\textsf{nonPositiveInteger, nonNegativeInteger, positiveInteger, } \\
&\phantom{{}=1}\textsf{decimal, integer, negativeInteger}\} \notag \\
&\textsf{atomicNumericLike} = \{\textsf{float, double, long, int, short, byte}\} \\
&\textsf{atomicUnsignedLike} = \{\textsf{unsignedLong, unsignedInt, unsignedShort,} \\
&\phantom{{}=1}\textsf{unsignedByte}\} \notag \\
&\textsf{booleanLike} = \{\textsf{boolean}\} 
\end{align*}
}

There is also a distinguished $\top$ kind for the $\top$ datatype for upward closure.
Algorithm~\ref{alg:pref} ($pref$) compares pairs of minimally required datatypes, and if two datatypes are comparable with respect to $\leq_s$, the greater datatype is removed from the set.
The resulting set $R'$ is still an antichain of datatypes with respect to $\leq_{lex}$.

\begin{algorithm}
\caption{$pref$}
\label{alg:pref}
  \DontPrintSemicolon
  \SetKw{In}{in}
  \SetKw{And}{and}
  \SetKw{With}{with}
  \SetKwInOut{Input}{Input}
  \Input{lexical datatype system $(T, \phi, \sim_s, \leq_s)$\\datatypes $R \subseteq T$}
  \KwOut{preferred datatypes $R' \subseteq T$}
  \BlankLine
  $R' := R$ \;
  \For{$\tau, \tau'$ \In $R$ \And $\tau \neq \tau'$}{
		\lIf{$[\tau]_{\sim_s} <_s [\tau']_{\sim_s}$}{
			$R' := R' \; \setminus \; \{\tau'\}$
		}
  }
\end{algorithm}

\subsubsection{Datatyped Event Stream for Learning}

For learning, every text in a document event stream needs to be mapped to its minimally required datatypes:
\begin{align}
{minReq}(w) &= {pref}({minLex}(w)) \;\;\; \text{for string }w\\
{dtyped}(e) &= \begin{cases}
{minReq}(lab(e)) &  \text{if } characters\\
e & \text{for other events}
\end{cases}
\end{align}

The learner also needs to be able to aggregate minimally required datatypes from different text contents.
Let $v, w$ be to strings over Unicode.
The minimally required datatypes that accept both strings are:

\begin{equation}
\label{eq:minreq}
{minReq}(v, w) = \text{max}_{\leq_{lex}} \, {minReq}(v)\, \cup \, {minReq}(w)
\end{equation}

The max$_{\leq_{lex}}$ operation guarantees a nonempty antichain with respect to $\leq_{lex}$ that cover both strings.

\begin{example}
Let $S = \{$\texttt{1}, \texttt{0}, \texttt{true}, \texttt{33}$\}$ be Unicode strings, then $minReq(S) = \{\textsf{boolean}, \textsf{unsignedByte}\}$.
\end{example}

\subsection{The Incremental Learner}

A famous result by Gold~\cite{Gold1967} states that the language class of unrestricted regular expressions is not learnable in the limit from positive examples only.
This result translates to dXVPAs because modules characterize regular languages over types and datatypes.
The full language class of datatyped event streams expressible in dXVPAs can therefore not be learned from example documents only, and restrictions are necessary.
Two restrictions originating from schema complexity are considered:

\begin{itemize}
	\item \emph{Simplicity of regular expressions.} \ Bex et al.~\cite{Bex2004} have examined 202 DTDs and XSDs and conclude that the majority of regular expressions in practical schema productions are simple because types occur only a small number of times in expressions.
	
	\item \emph{Locality of typing contexts.} \ Martens et al.~\cite{Martens2006} have studied 819 DTDs and XSDs from the web and XML standards, and typing elements in 98\% is local, i.e., the type of an element only depends on its parent.
\end{itemize}

To capture simplicity, Bex et al.~\cite{Bex2010a} define the class of single-occurrence regular expressions (SOREs).
In a SORE, a symbol occurs at most once, and the majority of schema productions in the wild belong to this class.
SOREs generate a 2-testable regular language, and $k$-testable regular languages~\cite{Garcia1990} are known to be efficiently learnable from positive examples only.

A $k$-testable regular language is fully characterized by a finite set of allowed substrings of length $k$, and learning is collecting the substrings.
This can be done efficiently by constructing a prefix tree acceptor (PTA), i.e., a DFA that accepts exactly the examples, and \emph{naming the states} according to the string prefixes that lead to them.
Merging states whose names share the same $(k-1)$-length suffix then generalizes the automaton.
This can be done implicitly while constructing a PTA.
The proposed learner utilizes this idea by embedding typing information in state names that are derived from prefixes of datatyped event streams.

\subsubsection{Typing Mechanisms}

Typing can be thought of as a function that determines the type of an element from its name and other elements in the document~\cite{Murata2005,Martens2006}.

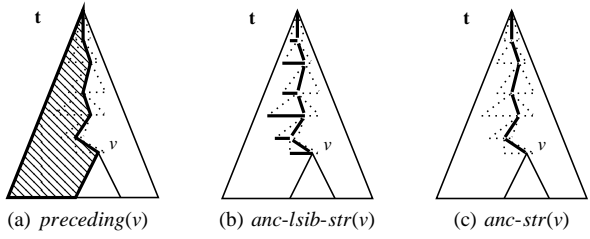
\begin{figure}
\centering
\tikzsetnextfilename{preceding-tree}
\subfloat[$preceding(v)$]{
	\label{fig:preceding_tree} {
	\usetikzlibrary{patterns}
\begin{tikzpicture}[semithick, inner sep=0pt, outer sep=0pt, font=\scriptsize]
\draw (0,0) -- (2,0) -- (1, 2.5)  node (v1) {}-- cycle;
\draw[dotted] (0.9,2.1) node (v2) {} -- (1,2.1) node (v3) {};
\draw[dotted] (0.8,1.8) node (v4) {} -- (1.1,1.8) node (v5) {};
\draw[dotted] (0.8,1.4) node (v6) {} -- (1,1.4) node (v7) {};
\draw[dotted] (0.6,1.1) node (v8) {} -- (1.1,1.1) node (v9) {};
\draw[dotted] (0.7,0.8) node (v10) {} -- (0.9, 0.8) node (v11) {};
\draw[dotted] (0.9,0.6) node (v12) {} -- (1.2, 0.6) node (v19) {};
\draw[dotted]  (v1) edge (v2);
\draw[dotted]  (v3) edge (v4);
\draw[dotted]  (v5) edge (v6);
\draw[dotted] (v7) edge (v8);
\draw[dotted]  (v9) edge (v10);
\draw[dotted]  (v11) edge (v12);
\draw (0.9,0) --(1.2, 0.6) -- (1.5, 0);
\node (v13) at (1.1,2.1) {};
\node (v14) at (1.2,1.8) {};
\node (v15) at (1.3,1.4) {};
\node (v16) at (1.3,1.1) {};
\node (v17) at (1,0.8) {};
\node (v18) at (1.3,0.6) {};
\draw[dotted]  (v1) edge (v13);
\draw[dotted]  (v13) edge (v3);
\draw[dotted]  (v3) edge (v14);
\draw[dotted]  (v14) edge (v5);
\draw[dotted]  (v5) edge (v15);
\draw[dotted]  (v15) edge (v7);
\draw[dotted]  (v7) edge (v16);
\draw[dotted]  (v16) edge (v9);
\draw[dotted]  (v9) edge (v17);
\draw[dotted]  (v17) edge (v11);
\draw[dotted]  (v11) edge (v18);
\draw[dotted]  (v18) edge (v19);
\fill[pattern=north west lines] (1, 2.5) -- (1,2.1) -- (1.1,1.8) -- (1,1.4) -- (1.1,1.1) -- (0.9, 0.8) -- (1.2, 0.6) -- (0.9,0) -- (0,0) -- cycle;
\draw[very thick] (1, 2.5) -- (1,2.1) -- (1.1,1.8) -- (1,1.4) -- (1.1,1.1) -- (0.9, 0.8) -- (1.2, 0.6) -- (0.9,0) -- (0,0) -- cycle;
\node at (0.4,2.4) {$\mathbf{t}$};
\node at (1.4,0.7) {$v$};
\node at (2.1,0.2) {};
\node at (-0.1,0.2) {};
\end{tikzpicture}
	}}
\tikzsetnextfilename{ancsib-tree}	
\subfloat[$anc\mbox{-}lsib\mbox{-}str(v)$]{
	\label{fig:ancsib_tree} {
	\begin{tikzpicture}[semithick, inner sep=0pt, font=\scriptsize]
\draw (0,0) -- (2,0) -- (1, 2.5) node (v1) {} -- cycle;
\draw (0.9,0) --(1.2, 0.6) -- (1.5, 0);
\draw[very thick] (0.9,2.1) node (v2) {} -- (1,2.1) node (v3) {};
\draw[very thick] (0.8,1.8) node (v4) {} -- (1.1,1.8) node (v5) {};
\draw[very thick] (0.8,1.4) node (v6) {} -- (1,1.4) node (v7) {};
\draw[very thick] (0.6,1.1) node (v8) {} -- (1.1,1.1) node (v9) {};
\draw[very thick] (0.7,0.8) node (v10) {} -- (0.9, 0.8) node (v11) {};
\draw[very thick] (0.9,0.6) node (v12) {} -- (1.2, 0.6) node (v19) {};
% \draw[dotted] (1, 2.5) -- (0.9,2.1);
% \draw[dotted]  (1,2.1) -- (0.8,1.8);
% \draw[dotted] (1.1,1.8) -- (0.8,1.4);
% \draw[dotted]  (1,1.4) -- (0.6,1.1);
% \draw[dotted]  (1.1,1.1) -- (0.7,0.8);
% \draw[dotted]  (0.9, 0.8) -- (0.9,0.6);
%\draw[dotted] (1, 2.5) node (v1) {} -- (0.9,2.1) -- (1,2.1) -- (1,1.4) -- (1.1,1.1) -- (0.9, 0.8) -- (1.2, 0.6);
\node at (0.4,2.4) {$\mathbf{t}$};
\node at (1.4,0.7) {$v$};
\node at (2.1,1) {};
\node at (-0.1,1) {};
%\node[draw, shape=circle, inner sep=0pt, minimum size=1pt, fill] at (1, 2.5) {};

\draw[dotted]  (v1) edge (v2);
\draw[dotted]  (v3) edge (v4);
\draw[dotted]  (v5) edge (v6);
\draw[dotted] (v7) edge (v8);
\draw[dotted]  (v9) edge (v10);
\draw[dotted]  (v11) edge (v12);
\node (v13) at (1.1,2.1) {};
\node (v14) at (1.2,1.8) {};
\node (v15) at (1.3,1.4) {};
\node (v16) at (1.3,1.1) {};
\node (v17) at (1,0.8) {};
\node (v18) at (1.3,0.6) {};
\draw[dotted]  (v1) edge (v13);
\draw[dotted]  (v13) edge (v3);
\draw[dotted]  (v3) edge (v14);
\draw[dotted]  (v14) edge (v5);
\draw[dotted]  (v5) edge (v15);
\draw[dotted]  (v15) edge (v7);
\draw[dotted]  (v7) edge (v16);
\draw[dotted]  (v16) edge (v9);
\draw[dotted]  (v9) edge (v17);
\draw[dotted]  (v17) edge (v11);
\draw[dotted]  (v11) edge (v18);
\draw[dotted]  (v18) edge (v19);
\draw[very thick]  (v1) -- (v3)-- (v5)  -- (v7) -- (v9) -- (v11) -- (v19);
\end{tikzpicture}
	}}
\tikzsetnextfilename{anc-tree}
\subfloat[$anc\mbox{-}str(v)$]{
	\label{fig:anc_tree} {
	\begin{tikzpicture}[semithick, inner sep=0pt, outer sep=0pt, font=\scriptsize]
\draw (0,0) -- (2,0) -- (1, 2.5) node (v1) {} -- cycle;
\draw (0.9,0) --(1.2, 0.6) -- (1.5, 0);
\draw[dotted] (0.9,2.1) node (v2) {} -- (1,2.1) node (v3) {};
\draw[dotted] (0.8,1.8) node (v4) {} -- (1.1,1.8) node (v5) {};
\draw[dotted] (0.8,1.4) node (v6) {} -- (1,1.4) node (v7) {};
\draw[dotted] (0.6,1.1) node (v8) {} -- (1.1,1.1) node (v9) {};
\draw[dotted] (0.7,0.8) node (v10) {} -- (0.9, 0.8) node (v11) {};
\draw[dotted] (0.9,0.6) node (v12) {} -- (1.2, 0.6) node (v19) {};
% \draw[dotted] (1, 2.5) -- (0.9,2.1);
% \draw[dotted]  (1,2.1) -- (0.8,1.8);
% \draw[dotted] (1.1,1.8) -- (0.8,1.4);
% \draw[dotted]  (1,1.4) -- (0.6,1.1);
% \draw[dotted]  (1.1,1.1) -- (0.7,0.8);
% \draw[dotted]  (0.9, 0.8) -- (0.9,0.6);
%\draw[dotted] (1, 2.5) node (v1) {} -- (0.9,2.1) -- (1,2.1) -- (1,1.4) -- (1.1,1.1) -- (0.9, 0.8) -- (1.2, 0.6);
\node at (0.4,2.4) {$\mathbf{t}$};
\node at (1.4,0.7) {$v$};
\node at (2.1,0.2) {};
\node at (-0.1,0.3) {};
%\node[draw, shape=circle, inner sep=0pt, minimum size=1pt, fill] at (1, 2.5) {};

\draw[dotted]  (v1) edge (v2);
\draw[dotted]  (v3) edge (v4);
\draw[dotted]  (v5) edge (v6);
\draw[dotted] (v7) edge (v8);
\draw[dotted]  (v9) edge (v10);
\draw[dotted]  (v11) edge (v12);
\node (v13) at (1.1,2.1) {};
\node (v14) at (1.2,1.8) {};
\node (v15) at (1.3,1.4) {};
\node (v16) at (1.3,1.1) {};
\node (v17) at (1,0.8) {};
\node (v18) at (1.3,0.6) {};
\draw[dotted]  (v1) edge (v13);
\draw[dotted]  (v13) edge (v3);
\draw[dotted]  (v3) edge (v14);
\draw[dotted]  (v14) edge (v5);
\draw[dotted]  (v5) edge (v15);
\draw[dotted]  (v15) edge (v7);
\draw[dotted]  (v7) edge (v16);
\draw[dotted]  (v16) edge (v9);
\draw[dotted]  (v9) edge (v17);
\draw[dotted]  (v17) edge (v11);
\draw[dotted]  (v11) edge (v18);
\draw[dotted]  (v18) edge (v19);
\draw[very thick]  (v1) -- (v3)-- (v5)  -- (v7) -- (v9) -- (v11) -- (v19);
\end{tikzpicture}
	}}
\caption{Typing of an element}
\label{fig:ancsib}
\end{figure}

Figure~\ref{fig:ancsib} illustrates typing mechanisms by representing the infoset of a document without text contents and identity constraints as a tree, where $lab(v)$ is the qualified element name of node $v$.
Efficient stream processing requires deterministic typing, and Martens et al.~\cite{Martens2006} therefore define \emph{1-pass preorder typing}~(1PPT): a schema allows 1PPT if the type of every node $v$ can be determined from the $preceding(v)$ subtree as shown in Figure~\ref{fig:preceding_tree}.
The authors surprisingly show that typing based on the ancestor-sibling string $anc\mbox{-}lsib\mbox{-}str(v)$ is sufficient for the 1PPT property.

Let $lsib(v) = lab(u_1) \cdots lab(u_m)\cdot lab(v)$ be a left-sibling string, where $u_1, \dots, u_m$ are the left siblings of $v$.
The ancestor-sibling string is then $anc\mbox{-}lsib\mbox{-}str(v) = lsib(i_1) \# lsib(i_2) \# \cdots \# lsib(i_n)$ such that $i_1$ is the root node, $i_n = v$, and $i_{j+1}$ is a child of $i_j$.
An example is in Figure~\ref{fig:ancsib_tree}.

Element Declaration Consistency~(EDC) and Unique Particle Attribution~(UPA) are syntactic restrictions for productions in XSD to ensure deterministic typing.
These restrictions are tighter than necessary for the 1PPT property.
In XSD, a node $v$ is typed by the ancestor string $anc\mbox{-}str(v)$ as shown in Figure~\ref{fig:anc_tree}.
The ancestor string is defined as $anc\mbox{-}str(v) = lab(i_1)\cdot lab(i_2) \cdots lab(i_n)$, where $i_1$ is the root node, $i_n = v$, and $i_{j+1}$ is a child of $i_j$.

\subsubsection{Incremental Update}

Named states in Algorithm~\ref{alg:incwvpa} ($incWeightedVPA$) are the foundation for state merging.
The algorithm iterates over document event stream $w$ in a single pass and returns an updated VPA and counters.
In a run, the algorithm maintains a stack, collects element names, and for every event, a next state is derived from three \emph{state naming functions} with signatures $call: Q \times \Sigma \to Q$, $int: Q \to Q$, and $ret: Q \times Q \times \Sigma \to Q$.
A transition is then stored to connect the current with the next state.

The three functions utilize the discussed typing mechanisms, and two \emph{state naming schemes} are proposed.

\begin{algorithm}
\caption{$incWeightedVPA$}
\label{alg:incwvpa}
  \DontPrintSemicolon
  \SetKw{In}{in}
  \SetKw{And}{and}
  \SetKw{Let}{let}
  \SetKwInOut{Input}{Input}
  \Input{VPA $A = (Q, q_0, F, Q, \delta)$ over $\Sigma \uplus T \uplus \overline{\Sigma}$\\lexical datatype system $(T, \phi, \sim_s, \leq_s)$\\state naming functions $call$, $int$, $ret$\\counters $\omega_Q$, $\omega_F$, $\omega_\delta$\\document event stream $w$}
  \KwOut{updated VPA $A$ and counters $\omega_Q$, $\omega_F$, $\omega_\delta$}
  \BlankLine
  $s := \bot$ \tcp*{empty stack}
  $q := q_0$ \tcp*{current state}
  \For{$e$ \In $dtyped(w)$}{
	\Switch{$eventType(e)$}{
	  \Case{$startElement$}{
		$\Sigma := \Sigma \cup \{lab(e)\}$\;
	    $q' := call(q, lab(e))$\;
	    $\omega_Q(q') := \omega_Q(q') + 1$ \;
		$\delta^{call} := \delta^{call} \cup \{q \xrightarrow{lab(e)/q} q'\}$ \;
		$\omega_\delta(q \xrightarrow{lab(e)/q} q') \! := \omega_\delta(q \xrightarrow{lab(e)/q} q') + 1$\;
		$s := s \cdot q$ \;
		$q := q'$
	  }
	  \Case{$endElement$}{
	  	\Let $vp = s$ \tcp*{$p$ is top}
	    $q' := ret(q, p, lab(e))$\;
	    $\omega_Q(q') := \omega_Q(q') + 1$ \;
	    $\delta^{ret} := \delta^{ret} \cup \{q \xrightarrow{\overline{lab(e)}/p} q'\}$ \;
	    $\omega_\delta(q \xrightarrow{\overline{lab(e)}/p} q')\! := \omega_\delta(q \xrightarrow{\overline{lab(e)}/p} q')\! +\! 1$\;
	    $s := v$ \;
	    $q := q'$
	  }
	  \Case{$characters$}{
	    $q' := int(q)$\; 
	    $\omega_Q(q') := \omega_Q(q') + 1$ \;
		$\delta^{int} := \delta^{int} \cup \{q \xrightarrow{\tau} q' \mid \tau \in lab(e)\}$\;	    \lFor{$\tau \in lab(e)$}{$\omega_\delta(q \xrightarrow{\tau} q') := \omega_\delta(q \xrightarrow{\tau} q') + 1$}
	    $q := q'$
	  }
	}
  }
  $F := F \cup \{q\}$\;
  $\omega_F(q) := \omega_F(q) + 1$
\end{algorithm}

\begin{definition}[State naming schemes]
\label{def:naming}
A state is a pair $(u, v)$ of typing context $u$ and left-sibling string $v$.
Symbols $\#$ and $\$$ are a left-sibling separator and a placeholder for text.

\begin{itemize}
	\item \emph{Ancestor-based.} \ A state $(u, v) \in (\Sigma^* \times (\Sigma \cup \{\$\})^*)$ is a pair of ancestor string and left-sibling string.

	\item \emph{Ancestor-sibling-based.} \ A state $(u, v) \in ((\Sigma \cup \{\$, \#\})^* \times (\Sigma \cup \{\$\})^*)$ is a pair of ancestor-sibling string and left-sibling string.
\end{itemize}

Initially, the intermediate VPA has a single nonaccepting start state $(\epsilon, \epsilon)$, no transitions, and counters are set to zero.
Next states and transitions are created inductively from the start state, and counters are increased.
For learning within the XSD language class, states must be ancestor based.
Beyond XSD but within the 1PPT language class, states must be ancestor-sibling based.
\end{definition}

\subsubsection{Local State Merging}

Based on Definition~\ref{def:naming}, every prefix of every datatyped event stream can characterize a state.
When complete ancestor, ancestor-sibling, and left-sibling strings are returned by state naming functions in Algorithm $incWeightedVPA$, the resulting automaton would accept exactly the learned documents similar to a PTA for regular languages.
Generalization by state merging is then embedded in the state naming functions by returning equivalence classes of named states.

The distinguishing criterion is \emph{locality}: two states are equal if they share the same $l$-local typing context and $k$-local left siblings.
The refined naming functions are:

\begin{align}
&int_{k, l}(q) = (\pi_1(q), \sigma_k(\pi_2(q) \cdot \$)), \label{eq:merge_int} \\
&ret_{k, l}(q, p, e) = (\pi_1(p), \sigma_k(\pi_2(p) \cdot lab(e))), \\
&call^{as}_{k, l}(q, e) = (\sigma_l(\pi_1(q) \cdot lab(e)), \epsilon), \\
&call^{als}_{k, l}((r_1\# \cdots \#r_n, v), e) = \notag \\
&\phantom{{}=1}(r_{n-l+1}\#\cdots\#r_n \# \sigma_k(v \cdot lab(e)), \epsilon) \label{eq:merge_als_call}
\end{align}
%&ret^{als}_{k, l}(q, p, e) = (\pi_1(p), \sigma_k(\pi_2(p) \cdot lab(e))).\label{eq:merge_als_ret}

Ancestor- and ancestor-sibling-based naming schemes require a different $call$ function denoted by superscripts $as$ and $als$ respectively.
The suffix function $\sigma_i(w)$ returns the $i$-length suffix of sequence $w$, and $\pi_i(x)$ denotes the $i$th field of tuple~$x$.
For $characters$ events, $int_{k,l}$ is the same under both naming schemes; the typing context remains unchanged, and $\$$ is appended to the left siblings as a placeholder.
Using a placeholder for the next state is sound because of the mixed-content restrictions in Definition~\ref{def:mixed}.
For $endElement$ events, $ret_{k, l}$ is also the same under both naming schemes; the next state inherits the typing context from stack state $p$, and a new left sibling is added to the ones in $p$. 
In case of a $startElement$ event, a new typing context is created, and left siblings are set to empty.

Parameters $k$ and $l$ specify the hypothesis space of the learner.
For the lower bound $k = l = 1$, the learnable language class is a strict subclass of DTD.
For $l \geq 1, k = 1$, both state naming schemes produce congruent automata, and the learnable language class is a strict subclass of XSD.
Greater parameters increase the learnable language class, but also the state space grows, and more examples are necessary for convergence.
If the true language class is not $k$-$l$-local or when parameters are chosen too small, an approximation is learned.

\subsubsection{Generating a dXVPA}

The intermediate VPA and its counters still need to be translated into a dXVPA.
Algorithm~\ref{alg:trim} ($trim$) creates a new intermediate VPA without zero-weight states and transitions.
Furthermore, $trim$ ensures a correct antichain of datatypes for internal datatype transitions between two states.

\begin{algorithm}
\caption{$trim$}
\label{alg:trim}
  \DontPrintSemicolon
  \SetKw{In}{in}
  \SetKw{And}{and}
  \SetKw{Let}{let}
  \SetKwInOut{Input}{Input}
  \Input{VPA $A = (Q, q_0, F, Q, \delta)$ over $\Sigma \uplus T \uplus \overline{\Sigma}$\\lexical datatype system $(T, \phi, \sim_s, \leq_s)$\\counters $\omega_Q$, $\omega_F$, $\omega_\delta$}
  \KwOut{VPA $A' = (Q', q_0, F', Q', \delta')$}
  \BlankLine
  $\delta^{'call} := \delta^{call} \setminus \{q \xrightarrow{c/q} q' \mid \omega_\delta(q \xrightarrow{c/q} q') = 0\}$ \;
  $\delta^{'ret} := \delta^{ret} \setminus \{q \xrightarrow{\overline{c}/q} q' \mid \omega_\delta(q \xrightarrow{\overline{c}/q} q') = 0\}$ \;
  $\delta^{'int} := \delta^{int} \setminus \{q \xrightarrow{\tau} q' \mid \omega_\delta(q \xrightarrow{\tau} q') = 0\}$ \;
  $\delta^{''int} := \emptyset$ \;
  \ForEach{$\{(q, q') \mid \exists \tau. q \xrightarrow{\tau} q' \in \delta^{'int}\}$}{
    \Let $R = \{\tau \mid q \xrightarrow{\tau} q' \in \delta^{'int}\}$\;
    $\delta^{''int} := \delta^{''int} \cup \{q \xrightarrow{\tau} q' \mid \tau \in \text{max}_{\leq_{lex}} R\}$\;
  }
  $\delta' = \delta^{'call} \uplus \delta^{'ret} \uplus \delta^{''int}$ \;
  $Q' := Q \setminus \{q \mid \omega_Q(q) = 0\}$ \;
  $F' := F \setminus \{q \mid \omega_F(q) = 0\}$ \;
\end{algorithm}

\begin{algorithm}
\caption{$genXVPA$}
\label{alg:genxvpa}
  \DontPrintSemicolon
  \SetKw{In}{in}
  \SetKw{And}{and}
  \SetKw{Let}{let}
  \SetKwInOut{Input}{Input}
  \Input{VPA $A = (Q, q_0, F, Q, \delta)$ over $\Sigma \uplus T \uplus \overline{\Sigma}$\\lexical datatype system $(T, \phi, \sim_s, \leq_s)$}
  \KwOut{dXVPA $A'$ over $(\Sigma, M, \mu, T, \phi)$, where $A' = (\{Q_m, e_m, X_m, \delta_m\}_{m \in M}, m_0, X_{m_0})$}
  \BlankLine
  $M := \{u \mid (u, v) \in Q \text{ and } u \neq \epsilon\}$ \;
  $m_0 := u \quad \text{such that}\quad q_0 \xrightarrow{c/q_0} (u, \epsilon) \in \delta^{call}$ \;
  \For{$m \in M$}{
  	$Q_m := \{(u, v) \in Q \mid u = m\}$ \;
    $e_m := (m, \epsilon)$\;
  	$X_m := \{q \in Q_m \mid q \xrightarrow{\overline{c}/p} q' \in \delta^{ret} \}$ \;
  	$\delta_m^{call} := \{q \xrightarrow{c/q} q'  \in \delta^{call} \mid q \in Q_m\}$\;
  	$\delta_m^{int} := \{q \xrightarrow{\tau} q' \in \delta^{int}  \mid q, q' \in Q_m\}$\;
	$\delta_m^{ret} := \{q \xrightarrow{\overline{c}/p} q' \in \delta^{ret}  \mid q \in Q_m\}$\;	
	$\delta_m^{ret} := \delta_m^{ret} \cup \{q \xrightarrow{\overline{c}/p} q' \mid q \in X_m \mbox{ and } \exists q_m. q_m \xrightarrow{\overline{c}/p} q' \in \delta_m^{ret} \}$\label{algo:single_exit}\;
	$\delta_m = \delta_m^{call} \uplus \delta_m^{call} \uplus \delta_m^{call}$\;
	\lIf{$\exists q.q \xrightarrow{c/q} e_m \in \delta^{call}$}{$\mu(m) := c$}
  }
  $A' := minimize(A')$
\end{algorithm}

Algorithm~\ref{alg:genxvpa} ($genXVPA$) generates a valid dXVPA from a trimmed intermediate VPA.
States are partitioned into modules based on their typing context. 
The initial module $m_0$ is the one called from state $(\epsilon, \epsilon)$.
The $call$ function from state naming guarantees that the entry of module $m$ is always state $(m, \epsilon)$.
Return transitions are added to all module exit states to ensure the single-exit property (Line~\ref{algo:single_exit}).

Algorithm~\ref{alg:minxvpa} ($minimize$) merges congruent modules.
Kumar et al.~\cite{Kumar2007} have shown that XVPA modules can be translated to DFAs, and this construction is extended to dXVPA modules.
The algorithm compares modules $m$ and $n$, and if they are reachable by the same element name and have congruent DFAs, $n$ folds into $m$ by redirecting calls and returns to corresponding states in $m$.
The state bijection $\varphi$ follows from bisimulation of the DFAs, and after a fold, $minimize$ restarts until no fold occurs.

\begin{algorithm}
\caption{$minimize$}
\label{alg:minxvpa}
  \DontPrintSemicolon
  \SetKw{In}{in}
  \SetKw{And}{and}
  \SetKw{Let}{let}
  \SetKw{Break}{break}
  \SetKwInOut{Input}{Input}
  \Input{dXVPA $A$ over $(\Sigma, M, \mu, T, \phi)$, where $A = (\{Q_m, e_m, X_m, \delta_m\}_{m \in M}, m_0, X_{m_0})$}
  \KwOut{minimized dXVPA $A$}
  \BlankLine
  \While{$\exists m \exists n. m, n \in M \text{ and } m \neq n \text{ and }\mu(m) = \mu(n)\text{ and }DFA_m \simeq DFA_n$}{
    \Let $\varphi: Q_n \to Q_m$ \tcp*{from bisimulation}
    \For{$q_n \xrightarrow{\overline{c}/p_i} q_i \in \delta^{ret}_n$}{
    	$\delta^{call}_i := \delta^{call}_i \setminus \{p_i \xrightarrow{c/p_i} e_n\} \cup \{p_i \xrightarrow{c/p_i} e_m\}$\;
    	$\delta^{ret}_m := \delta^{ret}_m \cup \{x_m \xrightarrow{\overline{c}/p_i} q_i \mid x_m \in X_m\} $
    }
    \For{$q_n \xrightarrow{\overline{c}/q_n} e_i \in \delta^{call}_n$}{
    	$\delta^{'ret}_i = \emptyset$\;
    	\For{$q_i \xrightarrow{\overline{c}/p_j} q_j \in \delta^{ret}_i$}{
    		\lIf{$\!\!j\! =\! n\!$}{$\!\delta^{'ret}_i\!\! := \!\!\delta^{'ret}_i \!\cup\! \{q_i \xrightarrow{\overline{c}/\varphi(p_j)} \varphi(q_j)\}$}
    		\lElse{$\delta^{'ret}_i := \delta^{'ret}_i \cup \{q_i \xrightarrow{\overline{c}/p_j} q_j\}$}
    	}
    	$\delta^{ret}_i := \delta^{'ret}_i$\;
    }
    \lIf{$n = m_0$}{$m_0 := m$}
	$M := M \setminus \{n\}$ \tcp*{remove module $n$}
	$\mu(n) := \emptyset$ \;
  }
\end{algorithm}

\subsubsection{Learner Properties}

Algorithm~\ref{alg:inclearner} assembles the learner.
Incrementally updating an intermediate VPA prevents information loss from premature minimization of dXVPA modules.
For a lexical datatype system, three naming functions, and parameters $k$ and $l$, the incremental learner computes a dXVPA from document event stream $w$.
The equivalent cXVPA can then check acceptance.

\begin{algorithm}
\caption{Incremental learner}
\label{alg:inclearner}
  \DontPrintSemicolon
  \SetKw{In}{in}
  \SetKw{And}{and}
  \SetKw{Let}{let}
  \SetKw{Break}{break}
  \SetKwInOut{Input}{Input}
  \Input{persistent VPA $A$\\lexical datatype system $dts = (T, \phi, \sim_s, \leq_s)$\\persistent counters $\omega = (\omega_Q$, $\omega_F$, $\omega_\delta)$\\state naming $f =(int_{k,l}, call_{k,l}, ret_{k,l})$ with $k, l$\\document event stream $w$}
  \KwOut{dXVPA $A'$}
  \BlankLine
  initially, $A = (\{(\epsilon, \epsilon)\}, (\epsilon, \epsilon), \emptyset, \{(\epsilon, \epsilon)\}, \emptyset)$\;
  $A, \omega := incWeightedVPA(A, dts, f, \omega, w)$\;
  $A' := genXVPA(trim(A, dts, \omega))$
\end{algorithm}

\begin{theorem}
\label{thm:properties}
The learner is (1) incremental, (2) set-driven, (3) consistent, (4) conservative, (5) strong-monotonic, and identifies a subclass of 1PPT mixed-content XML.
\end{theorem}

Incremental learning follows from Algorithm~\ref{alg:inclearner}.
A set-driven learner follows from calling $incWeightedVPA$ repeatedly for a set of examples and generating the dXVPA after the last one.
Set-driven learning is insensitive to the order of presented examples, and this property follows from state naming and treating states and transitions as sets.
A learner is consistent if all learned examples are accepted, conservative if a current hypothesis is kept as long as no contradicting evidence is presented, and strong-monotonic if the language increases with every example~\cite{Angluin1980,Jantke1991}.
These properties follow from updating sets of states and transitions in the intermediate VPA using the state naming functions.
States and call and return transitions are never deleted, and new ones are only added when observed.
Also, an internal transition on datatype $\tau$ is only removed if a new transition on $\tau'$ is added, where $\tau'$ covers $\tau$.

A learned dXVPA is always deterministic because of the restriction to $k$-$l$-local 1PPT.
Checking acceptance using the corresponding cXVPA is therefore linear in the length of the document event stream.

\subsection{Anomaly Detection Refinements}

The learning process could be targeted by poisoning~\cite{Cretu2008}, and two operations for dealing with adversarial training data are proposed.

\begin{algorithm}
\caption{$unlearn$}
\label{alg:unlearn}
  \DontPrintSemicolon
  \SetKw{In}{in}
  \SetKw{And}{and}
  \SetKw{Let}{let}
  \SetKwInOut{Input}{Input}
  \Input{VPA $A = (Q, q_0, F, Q, \delta)$ over $\Sigma \uplus T \uplus \overline{\Sigma}$\\lexical datatype system $dts = (T, \phi, \sim_s, \leq_s)$\\counters $\omega_Q$, $\omega_F$, $\omega_\delta$\\document event stream $w$}
  \KwOut{updated VPA $A$ and counters $\omega_Q$, $\omega_F$, $\omega_\delta$}
  \BlankLine
  $s := \bot$ \tcp*{empty stack}
  $q := q_0$ \tcp*{current state}
  \For{$e$ \In $dtyped(w)$}{
	\Switch{$eventType(e)$}{
	  \Case{$startElement$}{
	    $q' := \delta^{call}(q, lab(e))$ \;
	    $\omega_Q(q') := \omega_Q(q') - 1$ \;
		$\omega_\delta(q \xrightarrow{lab(e)/q} q') := \omega_\delta(q \xrightarrow{lab(e)/q} q') - 1$\;
		$s := s \cdot q$ \;
		$q := q'$
	  }
	  \Case{$endElement$}{
	  	\Let $vp = s$ \tcp*{$p$ is top}
	  	$q' := \delta^{ret}(q, lab(e), p)$ \;
	    $\omega_Q(q') := \omega_Q(q') - 1$ \;
	    $\omega_\delta(q \xrightarrow{\overline{lab(e)}/p} q') := \omega_\delta(q \xrightarrow{\overline{lab(e)}/p} q') - 1$\;
	    $s := v$ \;
	    $q := q'$
	  }
	  \Case{$characters$}{
	    $q' := \delta^{int}(q, \tau)$ for some $\tau \in lab(e)$\;
	    $\omega_Q(q') := \omega_Q(q') - 1$ \;
	    \lFor{$\tau \in lab(e)$}{$\omega_\delta(q \xrightarrow{\tau} q') := \omega_\delta(q \xrightarrow{\tau} q') - 1$}
	    $q := q'$
	  }
	}
  }
  $\omega_F(q) := \omega_F(q) - 1$\;
  $A := trim(A, dts, \omega_Q, \omega_F, \omega_\delta)$
\end{algorithm}

We distinguish poisoning attacks that are uncovered at some later time and poisoning attacks that remain hidden but are statistically rare.
Therefore, \emph{unlearning} removes a once learned example from the intermediate VPA, and \emph{sanitization} trims low-frequent transitions and states.

Algorithm~\ref{alg:unlearn} ($unlearn$) simulates a run on the document event stream that needs to be forgotten, traverses the intermediate VPA, and decrements counters.
The document must have been learned before at an earlier time for the operation to be sound.

\begin{algorithm}
\caption{$sanitize$}
\label{alg:sanitize}
  \DontPrintSemicolon
  \SetKw{In}{in}
  \SetKw{And}{and}
  \SetKw{Let}{let}
  \SetKwInOut{Input}{Input}
  \SetNoFillComment
  \Input{VPA $A = (Q, q_0, F, Q, \delta)$ over $\Sigma \uplus T \uplus \overline{\Sigma}$\\lexical datatype system $dts = (T, \phi, \sim_s, \leq_s)$\\counters $\omega_Q$, $\omega_F$, $\omega_\delta$}
  \KwOut{updated VPA $A'$ and counters $\omega'_Q$, $\omega'_F$, $\omega'_\delta$}
  \BlankLine
  \lFor{any defined transition $x$}{$\omega'_\delta(x) := \omega_\delta(x) - 1$}
  \For{$q \in Q$}{
  	$\omega'_Q(q) := \sum_{\text{transition } x \text{ to }q} \omega'_{\delta}(x)$ \;
  	\lIf{$q \in F$}{$\omega'_F(q) := \omega'_Q(q)$}
  }
  $A' := trim(A, dts, \omega'_Q, \omega'_F, \omega'_\delta)$\;
  \Let $Q_u$ be the unreachable states in $A'$\;
  \If{$Q_u \neq \emptyset$} {
	  \eIf(\tcp*[f]{revert changes}){$(F' \setminus Q_u) = \emptyset$}{
	    $\omega'_Q := \omega_Q$;\quad $\omega'_F := \omega_F$;\quad $\omega'_\delta := \omega_\delta$; \quad $A' := A$\;
	  }(\tcp*[f]{remove unreachable states})
	  {
	  	\For{$q \in Q_u$}{
  			\lFor{any transition $x$ to $q$}{$\omega'_{\delta}(x) := 0$}
  			$\omega'_Q(q) := \omega'_F(q) := 0$
  		}
	  	$A' := trim(A, dts, \omega'_Q, \omega'_F, \omega'_\delta)$
	}
  }
\end{algorithm}

Algorithm~\ref{alg:sanitize} ($sanitize$) trims low frequent states and transitions by decrementing all counters.
The algorithm has two stages.
First, counters for all transitions are decremented, and counters of states are recomputed.
Second, unreachable states are identified and decremented to zero for deletion.
If no final state is reachable, all weight counters are restored because sanitization is not applicable.

It should be stressed that sanitization should only be applied after a large number of examples have been learned.
The operation violates the consistent, conservative, and strong-monotonicity properties of the learner.
Also, after a $sanitize$ operation, $unlearn$ becomes unsound.

\begin{table*}
\newcommand*\rot[1]{\rotatebox{40}{\rlap{#1}}}
\scriptsize
\centering
\caption{Evaluation datasets}
\label{tab:datasets}
\begin{tabular}{l|l|l|l|lllllllllp{1cm}}
\hline\noalign{\smallskip}
                  & {Training} & \multicolumn{2}{c|}{{Testing}} &  \multicolumn{10}{c}{\rule{0pt}{1.05cm}}  \\
Dataset          & Normal & Normal &  Attack & \rot{XML tampering} & \rot{High node count} & \rot{Coercive parsing} & \rot{Script injection} & \rot{Command injection} & \rot{SQL injection} & \rot{SSRF attributes} & \rot{XML injection} & \rot{Signature wrapping}&  \\
\noalign{\smallskip}\hline\hline\noalign{\smallskip}
Carsale           & 50                      & 1000                    & 17                       & 1             & 3                       & 2                & 3                      & 2                       & 2             & 1    & 3             & 0                 &  \\
Catalog           & 100                     & 2000                    & 17                       & 1             & 3                       & 2                & 3                      & 2                       & 2             & 2    & 2             & 0                 &  \\
\noalign{\smallskip}\hline\noalign{\smallskip}
VulnShopOrder     & 200                     & 2000                    & 28                       & 2             & 4                       & 2                & 5                      & 3                       & 5             & 3    & 4             & 0                 &  \\
VulnShopAuthOrder & 200                     & 2000                    & 78                       & 0             & 0                       & 0                & 0                      & 0                       & 0             & 0    & 0             & 78    &  \\           
\noalign{\smallskip}\hline
\end{tabular}
\end{table*}

\section{Experimental Evaluation}

The proposed approach has been implemented in Scala~2.11.7, and two aspects of performance are considered: detection performance and learning progress.

\subsection{Measures}

By assuming binary classification between \emph{normal} and \emph{attack}, the following performance measures are computed from labeled datasets: recall/detection rate ($Re$), false-positive rate ($FPR$), precision ($Pr$), and $F_1$ for overall performance~\cite{Davis2006}.
Identification in the limit has a convergence point, but practical convergence can only be estimated by counting mind changes between incremental steps~\cite{DelaHiguera2010}.

\begin{definition}[Mind changes]
Mind changes $MC_{i}$ are the number of states and transitions whose counters switched from zero to one after learning document event stream $w_i$.
\end{definition}

Parameters $k$ and $l$ embody a strong combinatorial upper bound on the number of states and transitions for a finite number of elements.
In the worst case of randomness, convergence is reached when the state space is fully saturated.

\subsection{Datasets}

Table~\ref{tab:datasets} summarizes the four datasets.
The learner infers a dXVPA from training data, and performance is measured by validating the testing data with the corresponding cXVPA.
Datasets Carsale and Catalog have been synthetically generated using the stochastic XML generator ToXGene~\cite{Barbosa2002}.
For providing a realistic setting, a \emph{VulnShopService} and a randomized \emph{VulnShopClient} have been implemented for capturing SOAP messages.
This Apache~Axis2~1.6.0 SOAP/WS-* web service uses Apache~Rampart~1.6.0 for WS-Security and provides two service operations: regular shop orders (dataset VulnShopOrder) and digitally signed shop orders (dataset VulnShopAuthOrder).
For realism, the implementation strictly followed the Axis2 and Rampart examples.
The business logic utilized Java beans, and Java2WSDL automatically generated an Axis2 service from beans.
Names for operations and Java classes have been deliberately chosen to require types in a schema.

Attacks in synthetic datasets were added manually.
Attacks in the simulated datasets are recordings of actual attacks, e.g., WS-Attacker-1.7~\cite{ws-attacker} for Denial-of-Service (high node count, coercive parsing) and signature wrapping.

\subsection{Performance}

\subsubsection{Baseline Performance}

Schema validation using Apache~Xerces~2.9.1 established a baseline, and results are listed in Table~\ref{tab:baseline}.
The schemas for the Carsale and Catalog datasets were extracted from ToXGene configurations, and simple types were set to datatype \textsf{string} or more informative datatypes when applicable.
The VulnShopOrder and VulnShopAuthOrder datasets needed a schema collection from the web service because of the composed WS-* standards.

The schemas in synthetic datasets are free from extension points, and schema validation achieved good performance as expected.
The baseline for the simulated \emph{VulnShopService} however illustrated the effect of extension points.
Half of the attacks in VulnShopOrder were identified because of structural violations or datatype mismatches, but all Denial-of-Service attacks at extension points passed.
Furthermore, no signature wrapping attack was identified.

\subsubsection{Detection Performance}

Table~\ref{tab:detection_performance} summarizes the best results by the proposed algorithms for lowest parameters $k$ and $l$.
The best parameters were found in a grid search over values $k, l \in \{1, \dots, 5\}$ and the two naming schemes.

The proposed language-based anomaly detection approach outperformed the baseline.
No false positives were detected, and the best results were already achieved with the simplest parameters, i.e., ancestor-based state naming and $k=l=1$.
All structural anomalies caused by attacks were detected.
It should be stressed that $k=l=1$ was a good-enough approximation of the language to identify attacks, but more sound types were inferred for $l>1$.

Some script and command injection attacks were not identified.
These attacks have in common that exploitation code appears in texts and use CDATA fields to hide special characters, e.g., angled brackets and ampersands, from the XML parser's lexical analysis.
The lexical datatype system is too coarse in this case because the inferred datatype \textsf{normalizedString} permits the attack-identifying characters.

\begin{table}
\centering
\scriptsize
\caption{Schema validation baseline performance}
\label{tab:baseline}
\begin{tabular}{l|llll}
\hline\noalign{\smallskip}
%                 & \multicolumn{4}{l}{Baseline detection performance} \\
Dataset          & $Pr$       & $Re$        & $FPR$     & $F_1$       \\
\noalign{\smallskip}\hline\hline\noalign{\smallskip}
Carsale          & 100\%      & 82.35\%     & 0\%       & 90.32\%     \\
Catalog          & 100\%      & 76.47\%     & 0\%       & 86.67\%     \\
\noalign{\smallskip}\hline\noalign{\smallskip}
VulnShopOrder    & 100\%      & 50\%        & 0\%       & 66.67\%     \\
VulnShopAuthOder & undef.        & 0\%         & 0\%      & undef.      \\
\noalign{\smallskip}\hline  
\end{tabular}
\end{table}

\begin{table}
\centering
\scriptsize
\caption{Best performance using ancestor-based states}
\label{tab:detection_performance}
\begin{tabular}{l|l|llll}
\hline\noalign{\smallskip}
Dataset          &   $k\; l$    & $Pr$     & $Re$     & $FPR$    & $F_1$    \\
\noalign{\smallskip}\hline\hline\noalign{\smallskip}
Carsale                       & 1       1      & 100\%    & 100\%    & 0\%      & 100\%    \\
Catalog                & 1       1      & 100\%    & 82.35\%    & 0\%      & 90.32\%    \\
\noalign{\smallskip}\hline\noalign{\smallskip}
VulnShopOrder                & 1       1      & 100\%    & 92.86\%    & 0\%      & 96.30\%    \\
VulnShopAuthOrder       & 1       1      & 100\%    & 100\%    & 0\%      & 100\%   \\
\noalign{\smallskip}\hline
\end{tabular}
\end{table}

\subsubsection{Learning Progress}

\begin{figure*}
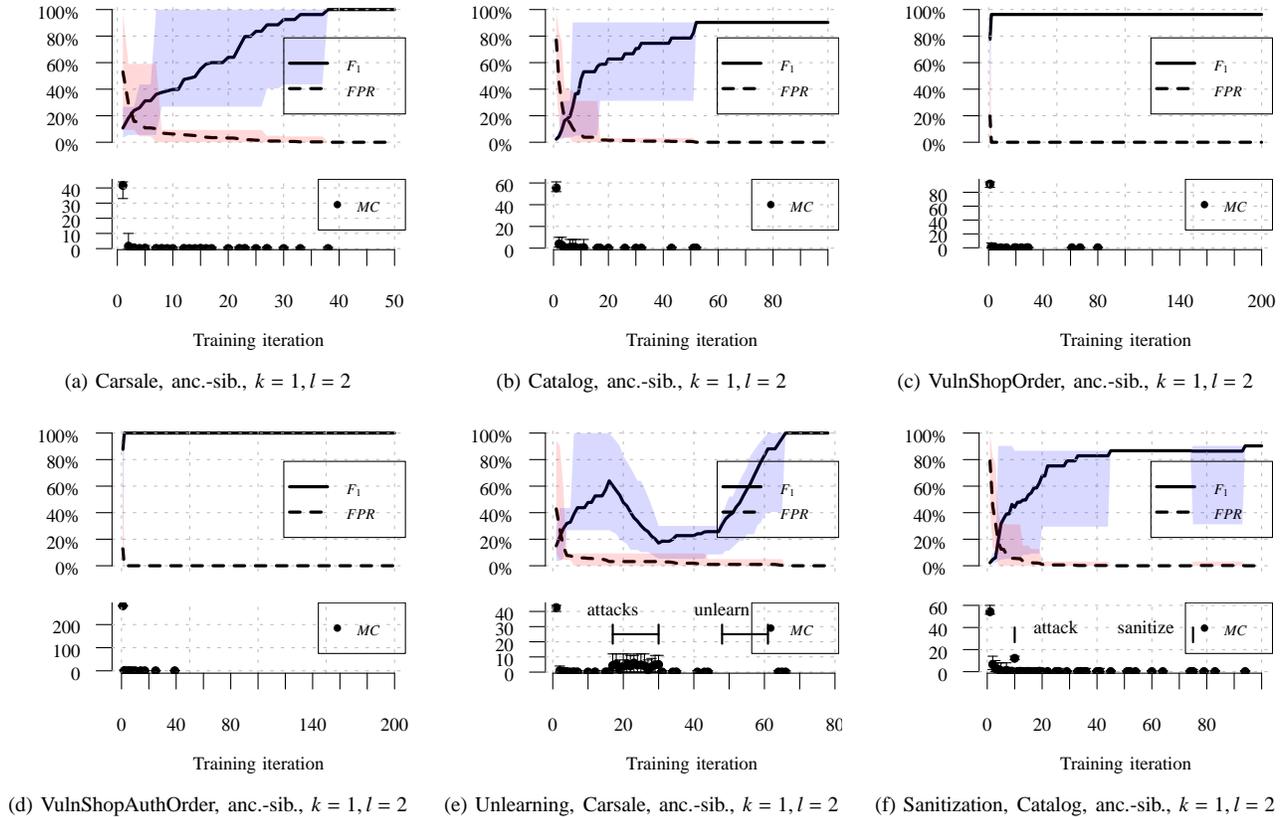

\scriptsize
	\tikzsetnextfilename{plot-carsale-ancsib-12}
	\subfloat[Carsale, anc.-sib., $k=1, l=2$]{
		\label{fig:highlights_carsale}{
		\input{R/learningProgress/carsale_ancsib_k1_l2.tikz}
	}}
	\tikzsetnextfilename{plot-catalog-ancsib-12}
	\subfloat[Catalog, anc.-sib., $k=1, l=2$]{
		\label{fig:highlights_catalog}{
		\input{R/learningProgress/catalog_ancsib_k1_l2.tikz}
	}}
	\tikzsetnextfilename{plot-vulnshoporder-12}
	\subfloat[VulnShopOrder, anc.-sib., $k=1, l=2$]{
		\label{fig:highlights_vulnshoporder}{
		\input{R/learningProgress/VulnShopOrder_ancsib_k1_l2.tikz}
	}}

	\tikzsetnextfilename{plot-vulnshopauthorder-12}
	\subfloat[VulnShopAuthOrder, anc.-sib., $k=1, l=2$]{
		\label{fig:highlights_vulnshopauthorder}{
		\input{R/learningProgress/VulnShopAuthOrder_ancsib_k1_l2.tikz}
	}}
	\tikzsetnextfilename{plot-unlearn-carsale-ancsib-12}
	\subfloat[Unlearning, Carsale, anc.-sib., $k=1, l=2$]{
		\label{fig:unlearning_carsale_ancsib_12}{
		\input{R/learningProgress/carsale_unlearn_ancsib_k1_l2.tikz}
	}}
	\tikzsetnextfilename{plot-sanitize-catalog-ancsib-12}
	\subfloat[Sanitization, Catalog, anc.-sib., $k=1, l=2$]{
		\label{fig:sanitize_catalog_ancsib_12}{
		\input{R/learningProgress/catalog_sanitize_ancsib_k1_l2.tikz}
	}}
	\caption{Learning progress}
	\label{fig:highlights}
\end{figure*}

Learning progress was measured in mind changes, and Figures~\ref{fig:highlights_carsale}--\ref{fig:highlights_vulnshopauthorder} summarize the fastest converging settings for the four datasets.
When converged, the performance coincided with Table~\ref{tab:detection_performance}.
In every training iteration, the learner randomly drew a training document without replacement for learning, and the validator checked acceptance of testing data for measuring improvements.
Because of randomness, runs were repeated 15 times, average values for $F_1$ and $FPR$ were computed, and the error regions in the plots illustrate minimal and maximal values in the random learning processes.

The first training example always caused many mind changes because there were no states and transitions yet.
The strong-monotonicity property guarantees that detection performance either increases or stays the same after learning an example and assuming it is not a poisoning attack.

In the real world, detection performance is not observable but mind changes are.
As shown in the figures, mind changes became less frequent over time, and a long period of zero mind changes could be a heuristic for convergence.

The quick convergence in Figure~\ref{fig:highlights_vulnshoporder} and~\ref{fig:highlights_vulnshopauthorder} stemmed from the simplicity of the language automatically generated by Java2WSDL.
The generator only supports sequential (member variables) and iterating (arrays) productions but no choice.
A few examples were sufficient for finding a good-enough approximation with small parameters $k$ and $l$.

\subsubsection{Unlearning and Sanitization}

Unlearning reverses learning, and Figure~\ref{fig:unlearning_carsale_ancsib_12} illustrates the effects.
In this scenario, a successful attacker was able to feed poisoning attacks to the learner, and performance dropped accordingly.
At a later time, a hypothetical expert identified the poisoning attacks and started unlearning them.
The detection performance recovered, and knowledge gained in between attacks and unlearning remained in the model.

Sanitization trims low-frequent states and transitions.
A single hidden poisoning attack was injected after 10\% learning progress, and there was an impact on performance.
After 75\% progress, sanitization was performed.
Figure~\ref{fig:sanitize_catalog_ancsib_12} shows the effects of sanitization.
In at least one of the 15 trials, the learner had no stable language representation at the moment of sanitization.
Good knowledge was trimmed, performance dropped, and more mind changes after sanitization were necessary to recover again.
Knowledge gained from a single example could be lost by sanitization.

\section{Related Work}

This work focuses on XML stream validation because of large documents and open-ended streams (e.g., XMPP).
Stream validation has been introduced by Segoufin and Vianu \cite{Segoufin2002} using finite-state machines and pushdown automata.
Kumar et al.~\cite{Kumar2007} consider document event streams as visibly pushdown languages (VPLs), a class of deterministic context-free languages, and the authors propose XVPAs as a better representation.
XVPAs have therefore been extended with datatypes for text contents.

Schema inference from a set of documents focuses on finding simple regular expressions for schema productions.
Beyond the expressiveness of DTD, Chidlovskii~\cite{Chidlovskii2001a} and Ml\'{y}nkov\'{a} and Ne\v{c}ask\'{y}~\cite{Mlynkova2013} propose grammar-based approaches, where infoset tree nodes turned into productions.
These productions are then generalized by determinism constraints~\cite{Chidlovskii2001a} and heuristics~\cite{Mlynkova2013}.
Bex et al.~\cite{Bex2007} propose schema inference in terms of tree automata, where up to $k$ ancestor elements in a document characterize a type.
This work has motivated the use of locality as a generalization strategy.
Lexical subsumption for datatype inference was fist mentioned by Chidlovskii~\cite{Chidlovskii2001a} and Hegewald et al.~\cite{Hegewald2006}; however, not all XSD datatypes have been considered.
The proposed approach considers a datatype choice instead of a single datatype, all distinguishable XSD datatypes are used, and a preference heuristic refines a choice.

With respect to anomaly detection, Menahem et al.~\cite{Menahem2015} propose a feature extraction process for documents, so existing machine-learning algorithms can be reused, but structural information is lost.
A schema is assumed to be available, and this direction has therefore not been further pursued.
Another anomaly detection approach specifically for tree structures is based on geometry.
Rieck~\cite{Rieck2009} introduces tree kernels as measures of shared information between two parse trees.
Kernels enable global and local anomaly detection, and this method could eventually be extended to XML infoset trees.
Global anomaly detection finds a volume-minimal sphere that encloses the vector-embedded trees, and local anomaly detection computes kernel-based distances to the nearest neighbors.
Approximate tree kernels~\cite{Rieck2010} are a trade-off for reducing computational costs.
However, this method assumes a tree which conflicts with streaming requirements.

\section{Conclusions}

This paper proposes a grammatical inference approach for learning the accepted language of an XML-based system.
Schema validation is ineffective as a defense mechanism when extension points are present.
For language-based anomaly detection, an automaton is inferred from examples, so documents with unexpected structure or text contents can be identified.
It is also possible to translate such an automaton into a schema~\cite{Kumar2007}.
The contributions are dXVPAs as language representations for mixed-content XML, cXVPAs as an optimization of dXVPAs for efficient stream validation, algorithms for datatype inference from text, an incremental learner, and an experimental evaluation in synthetic and realistic scenarios.

The dXVPAs capture well-nested event streams, i.e., linearizations of trees, but no integrity constraints, to stay within a language class that allows efficient stream validation.
This approach is nevertheless effective as a detection method because a learned language has no extension points.
Improving the learning setting from $k$-$l$-local languages toward more powerful ones, e.g., by query learning~\cite{DelaHiguera2010}, is a major open research question.
Inferring and validating integrity constraints are also open research questions; however, Arenas et al.~\cite{Arenas2014} have already shown that this problem is computationally much harder.

Simple parameters ($k=1, l=2$) for the learner outperformed baseline schema validation in experiments; nonetheless, there are limitations.
Some attacks in experiments could not be identified because lexical spaces of XSD datatypes are too coarse.
Introducing more fine-grained datatypes would improve the detection rate.
Also, repetitions are not bounded, and an order on unordered attributes is assumed.
Repetition bounds and unordered attributes are two additional open research questions.

Finally, the unlearning and sanitization operations help to deal with adversarial training data, but the operations only apply after a poisoning attack has happened.
The experiments indicated that the momentum of mind changes in the learning progress could be a heuristic for identifying a poisoning attack automatically while it is learned.

\ifCLASSOPTIONcompsoc
  \section*{Acknowledgments}
\else
  \section*{Acknowledgment}
\fi
The research has been supported by the Christian Doppler Society, and the results were produced while the author was affiliated with the Christian Doppler Laboratory for Client-Centric Cloud Computing, JKU Linz, Austria.


\bibliographystyle{IEEEtran}
\bibliography{IEEEabrv,literature}
\end{document}